\documentclass{article}

\usepackage[english]{babel}
\usepackage{ctex}
\usepackage{amssymb}
\usepackage{amsmath}
\usepackage{pdfpages}
\usepackage{caption}
\usepackage[letterpaper,top=2cm,bottom=2cm,left=3cm,right=3cm,marginparwidth=1.75cm]{geometry}

\usepackage{amsmath}
\usepackage{graphicx}
\usepackage[colorlinks=true, linkcolor=black, citecolor=blue]{hyperref}
\usepackage{lineno}
\usepackage{color}
\usepackage{xcolor}
\usepackage[superscript]{cite}
\usepackage{float}
\usepackage{mdframed} 
\usepackage{booktabs}
\usepackage[most]{tcolorbox}
\newcommand{\ourF}{DESRO}
\newcommand{\ourM}{DESRO}

\usepackage{graphicx} 
\usepackage{caption}  
\usepackage{multirow}
\usepackage{booktabs}
\newcounter{suppfigure}

\newcommand{\suppfigurecaption}[1]{
  \refstepcounter{suppfigure} 
}
\newcommand{\citep}[3]{#1, #2 \cite{#3}}

\makeatletter
\renewcommand{\maketitle}{\bgroup\setlength{\parindent}{0pt}
\begin{flushleft}
  \textbf{\@title}

  \@author
\end{flushleft}\egroup
}
\makeatother

\begin{document}
\title{Deciphering Scientific Reasoning Steps from Outcome Data for Molecule Optimization}
\author{Zequn Liu$^{1*}$, Kehan Wu$^{2*}$, Shufang Xie$^{1}$, Zekun Guo$^1$, Wei Zhang$^2$, Tao Qin$^1$, \\
Renhe Liu$^3$, Yingce Xia$^{1}$\\
$^1$ Zhongguancun Academy\\
$^2$ School of Computer Science and Technology, University of Science and Technology of China\\
$^3$ Global Health Drug Discovery Institute\\
$^*$ Equal contribution.\\
Emails: \texttt{liuzequn@bza.edu.cn}, \texttt{wu\_2018@mail.ustc.edu.cn}, \texttt{guozekun@bza.edu.cn}, \texttt{weizhang\_cs@mail.ustc.edu.cn}, \texttt{qintao@bza.edu.cn}, \texttt{renhe.liu@ghddi.org}, \texttt{xiayingce@bza.edu.cn}
}

\maketitle

\begin{abstract}
Scientific discovery typically requires rigorous logic and reasoning. While emerging reasoning models hold promise for automating this labor-intensive process, their training is hindered by a critical supervision gap: experimental outcomes are abundant, whereas intermediate reasoning steps are rarely documented at scale. 
To bridge this gap, we propose \ourF{}, a framework for \underline{de}ciphering \underline{s}cientific \underline{r}easoning from \underline{o}utcomes. 
Our insight is that latent reasoning leaves traces in outcomes. By analyzing shared patterns and key differences within grouped data, a large language model (LLM) can recover the underlying logic.
We instantiate this framework in molecule optimization, a pivotal stage in drug discovery that traditionally relies on the iterative reasoning of medicinal chemists. 
Across 2.3 million molecular property records, our framework infers optimization rationales by grouping molecules with shared fragments, then using an LLM to analyze how structural variations correlate with property differences. 
Based on the derived data, we train a model that conducts molecule optimization through an interpretable reasoning process. 
\ourM{} achieves the highest success rates on 15 out of 18 tasks, spanning both single- and multi-property optimization of bioactivity and ADMET properties. The reasoning process enables robust generalization to out-of-distribution scenarios, including novel property combinations, unseen biological targets, and unseen properties defined solely by natural language descriptions. In retrospective case studies under strict temporal splits, the model autonomously reconstructs expert-level lead optimization trajectories. Additionally, our framework extends beyond molecule optimization to reaction ligand selection. Our results establish deciphering reasoning steps from outcome data as a viable paradigm for enabling scientific reasoning, providing a scalable approach to accelerate scientific discovery.
\end{abstract}

\newpage

\section*{Introduction}
\noindent Scientific discovery fundamentally requires systematic reasoning and logical inference \cite{jiang2025deepseek,zheng2025large}. Traditionally, this reasoning has been carried out by scientists through deliberate analysis, hypothesis formation, and iterative verification \cite{dossetter2013matched,Gioiello2020}. Recent advances in reasoning models, which can decompose complex problems and plan intermediate steps, offer a promising path toward automating this labor-intensive process \cite{guo2025deepseek,jaech2024openai,yang2025qwen3,comanici2025gemini}. Yet bringing such paradigms into scientific scenarios remains constrained by a critical supervision gap for model training\cite{Zheng2025}. In many scientific domains such as chemistry and biology, experimental outcomes are abundant, while the intermediate reasoning steps that guided successive designs are far less frequently recorded and analyzed, preventing the models from learning the process of scientific reasoning\cite{wang2025medicalreasoningerallms,liu2025chemauharnessreasoningllms}.

This supervision gap is particularly evident in molecule optimization, a pivotal process throughout the drug discovery pipeline \cite{erikawa2021mermaid,ahmed2025ai,feng2025hierarchical,chakraborty2023utilizing}. It aims at refining chemical structures to improve diverse properties such as absorption, distribution, metabolism, excretion, and toxicity (ADMET) \cite{sliwoski2014computational,elton2019deep,du2024machine,fromer2023computer}.
Traditionally, optimization has relied on the iterative reasoning of medicinal chemists. However, this reasoning process is not comprehensively recorded. Scientific literature typically reports detailed design strategies only for a limited subset of representative successes. In contrast, large-scale repositories such as PubChem Bioassay \cite{kim2023pubchem} store molecules paired with assay readouts as isolated entries, without capturing the design context or the rationale linking successive modifications \cite{gaulton2012chembl,wishart2006drugbank}. As a result, many existing computational methods still simplify molecule optimization as direct generation \cite{jin2018junction,fu2022reinforced,dey-etal-2025-mathtt,zhu2025conditional,dey2025large,wu2024leveraging,ye2025drugassist,edwards2025mclm,liu2023multi}, producing optimized molecules without modeling the reasoning process.

To this end, we propose \ourF{}, a framework for \underline{de}ciphering \underline{s}cientific \underline{r}easoning from \underline{o}utcomes. Our key idea is that the unobserved reasoning trace can be recovered as a latent variable from groupwise contrasts in experimental outcomes. For a group of outcome records with different measurements, we extract their shared patterns and key differences, and infer the latent rationale behind outcome changes. Facilitated by the recent advances of large language models (LLMs) \cite{dubey2024llama,jaech2024openai,wei2022cot}, this reconstruction can be efficiently carried out at scale. We instantiate this framework in molecule optimization. We curate 2.3 million molecular property outcome records, primarily covering bioactivity and ADMET. For each target property, molecules are grouped by common fragments and decomposed into a common fragment and unique fragments, isolating the structural changes for property improvement. An LLM is employed to analyze fragment-level variations and their impact on properties, thereby deriving reasoning steps for molecule optimization. These reasoning steps establish a reasoning chain that links a starting molecule to its optimized counterpart via property analysis, rule induction, and targeted fragment modification. The derived data enables us to train a reasoning model for molecule optimization. By finetuning a general-purpose LLM on these reasoning traces, we enable the model to perform molecule optimization through an interpretable reasoning process rather than direct generation.


\ourM{} demonstrates superior performance on reasoning for molecule optimization tasks. First, in both single-property and multi-property optimization tasks involving ADMET and bioactivity, \ourM{} consistently outperforms general-purpose and domain-specific baselines by transforming direct generation into a structured reasoning procedure. Second, explicit reasoning improves generalization in molecule optimization. Through compositional reasoning, \ourM{} generalizes to novel combinations of multiple properties. By interpreting natural language descriptions, it successfully optimizes molecules for novel properties not encountered during training. Third, we demonstrate the real-world utility of \ourM{} through retrospective lead optimization studies under a strict temporal split on MALT1 and CBL-B. In these scenarios, \ourM{} successfully reconstructed expert-level reasoning trajectories and identified high-quality candidate molecules. Additionally, we show that \ourF{} extends beyond molecule optimization to another scientific design problem, reaction ligand selection. Collectively, these findings suggest that deciphering reasoning traces from outcome data offers a scalable and effective paradigm for constructing scientific reasoning models.

\section*{Results}
\subsection*{Deciphering reasoning steps and building a reasoning model for molecule optimization}

\begin{figure}[!htp]
    \centering
    \includegraphics[width=1.0\linewidth]{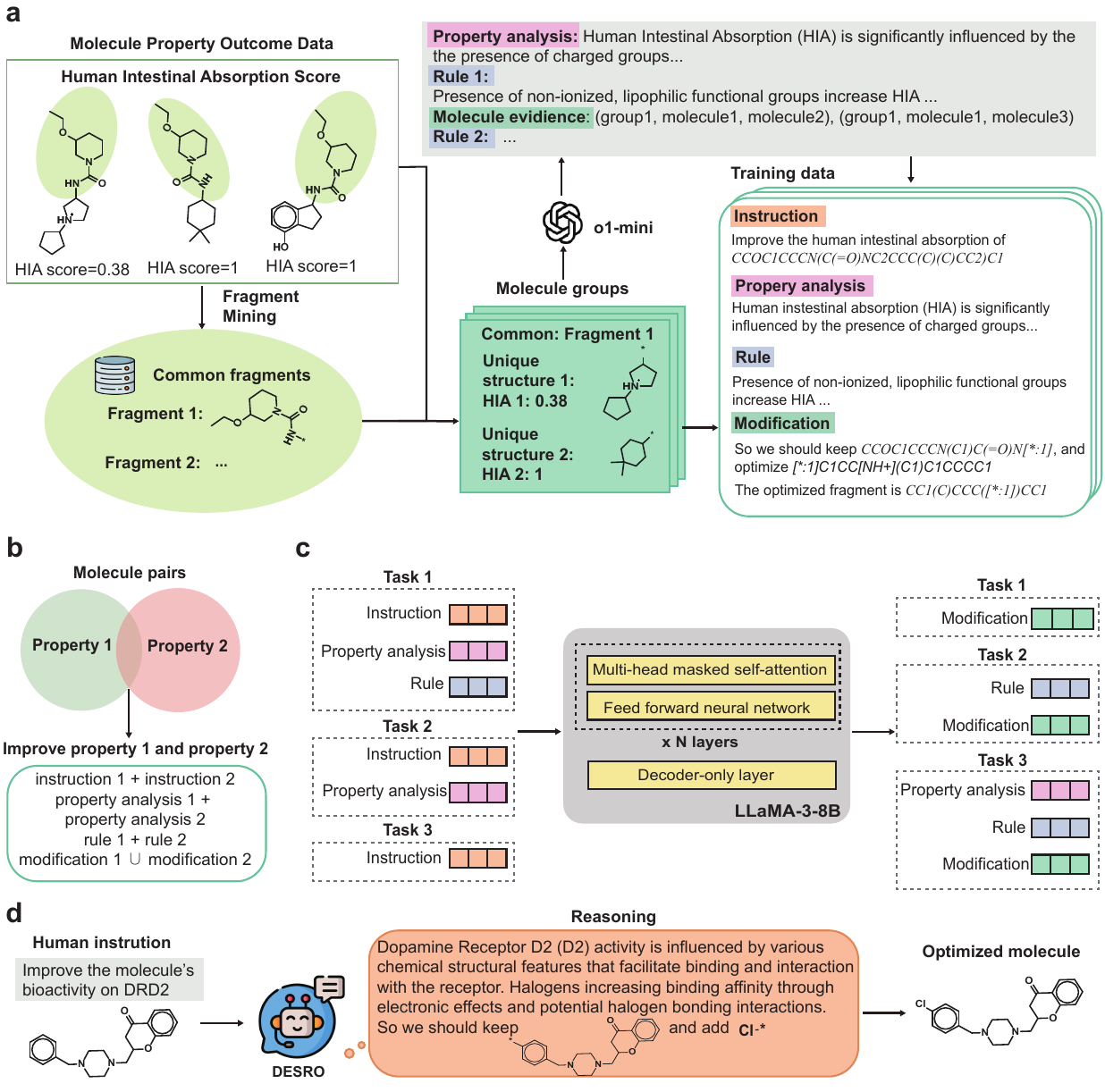} 
    \captionsetup{labelformat=empty}
    \caption{\textbf{Figure 1. Overview of \ourM{}. }
\textbf{a,} The pipeline of \ourF{} for molecule optimization. Large-scale molecular property datasets are processed using fragment mining to identify common fragments. o1-mini is then employed to induce reasoning steps, including property analyses and optimization rules, from these molecule groups, resulted in a reasoning-annotated training dataset.
\textbf{b,} Construction of multi-property optimization examples. Molecule pairs that appear across multiple property domains are identified, and their corresponding instructions, analyses, rules, and modifications are aggregated to enable joint multi-property optimization.
\textbf{c,} The multi-task training framework. A LLaMA-3-8B model is fine-tuned using a multi-task approach, where it learns to predict the remaining reasoning components and modification steps based on varied input configurations.
\textbf{d,} The inference workflow of the reasoning model. Given an initial molecule and a natural language optimization instruction, the model first generates an explicit reasoning trajectory, followed by the corresponding optimized molecular structure.}
    \label{fig:1}
\end{figure}

\noindent Although intermediate scientific reasoning is not frequently annotated, many datasets contain outcome records that implicitly reflect the decision logic behind iterative design. \ourF{} infers such latent logic by organizing related outcome records into groups, identifying what remains invariant across the group and what varies, and attributing outcome changes to these variations. Recent LLMs make this inference practical and scalable.

We instantiate this framework in molecule optimization (\textbf{Fig. \ref{fig:1}a}). We curated a dataset of 2.3 million molecules annotated with bioactivity and ADMET properties, organized by assay. For each assay, we apply a fragment mining algorithm \cite{gengnovo} to identify fragments common to multiple molecules, and group molecules by these shared fragments. Within each group, each molecule is partitioned into a common fragment and a set of unique fragments, which isolates the structural edits that correlate with property differences. 

To reconstruct optimization rationales at scale, we use o1-mini \cite{jaech2024openai} to interpret how variations in the unique fragments relate to measured property changes. 
We prompt o1-mini with structured inputs comprising the property definition and the grouped molecular data. Each group is represented by the shared fragment, the molecule-specific unique fragments, and the corresponding property values. The prompt instructs o1-mini to: (1) analyze the chemical principles about the target property, (2) compare unique structures within groups, focusing on molecule pairs with significant property differences, and (3) induce detailed chemical rules that relate structural features to property outcomes (\textbf{Supplementary Fig. \ref{fig:prompt}}). Based on this process, o1-mini outputs a general property analysis, a set of induced optimization rules, and identified supporting molecule pairs (i.e., two molecules with distinct property values whose fragment differences align with a specific rule). By mapping these molecule pairs to their associated fragment differences, property analyses, and optimization rules, we reconstructed a textual reasoning process for molecular property optimization.

This approach naturally extends from single-property to multi-property optimization (\textbf{Fig. \ref{fig:1}b}). By identifying molecule pairs that are shared across multiple properties, we can aggregate their individual property analyses and optimization rules. The corresponding modified fragments from each property are then combined to form composite fragments, enabling joint optimization across multiple properties. Finally, we obtained 197k distinct molecule optimization reasoning processes across both single-property optimization and multi-property optimization (\textbf{Supplementary Fig. \ref{fig:data_stat}}, \textbf{Supplementary Fig. \ref{fig:example}}).

We fine-tuned a LLaMA-3-8B \cite{dubey2024llama} model using supervised fine-tuning (SFT) on the collected reasoning data. To support flexible input formats, we adopted a multi-task learning framework (\textbf{Fig. \ref{fig:1}c}). During training, the model was fine-tuned on three input configurations: (1) instruction only, (2) instruction with property analysis, and (3) instruction with property analysis and rule, with the objective of predicting the remaining components in each case. 

The resulting model serves as a versatile molecule optimization reasoning model capable of generating interpretable optimization steps. Given the SMILES of a molecule and a textual optimization instruction such as \textit{improve Human Intestinal Absorption (HIA)}, the model outputs an optimized molecule together with an explicit reasoning process (\textbf{Fig. \ref{fig:1}d}). It first performs a property analysis summarizing the property definition and its principal structural and physicochemical determinants. It then derives an optimization rule conditioned on the analysis and the input structure. Finally, it edits the molecule by preserving a selected fragment and replacing another fragment according to the rule, producing the SMILES of the optimized molecule.

\subsection*{Accurate molecule optimization across diverse properties}
\begin{figure}[!htp]
    \centering
    \includegraphics[width=1.0\linewidth]{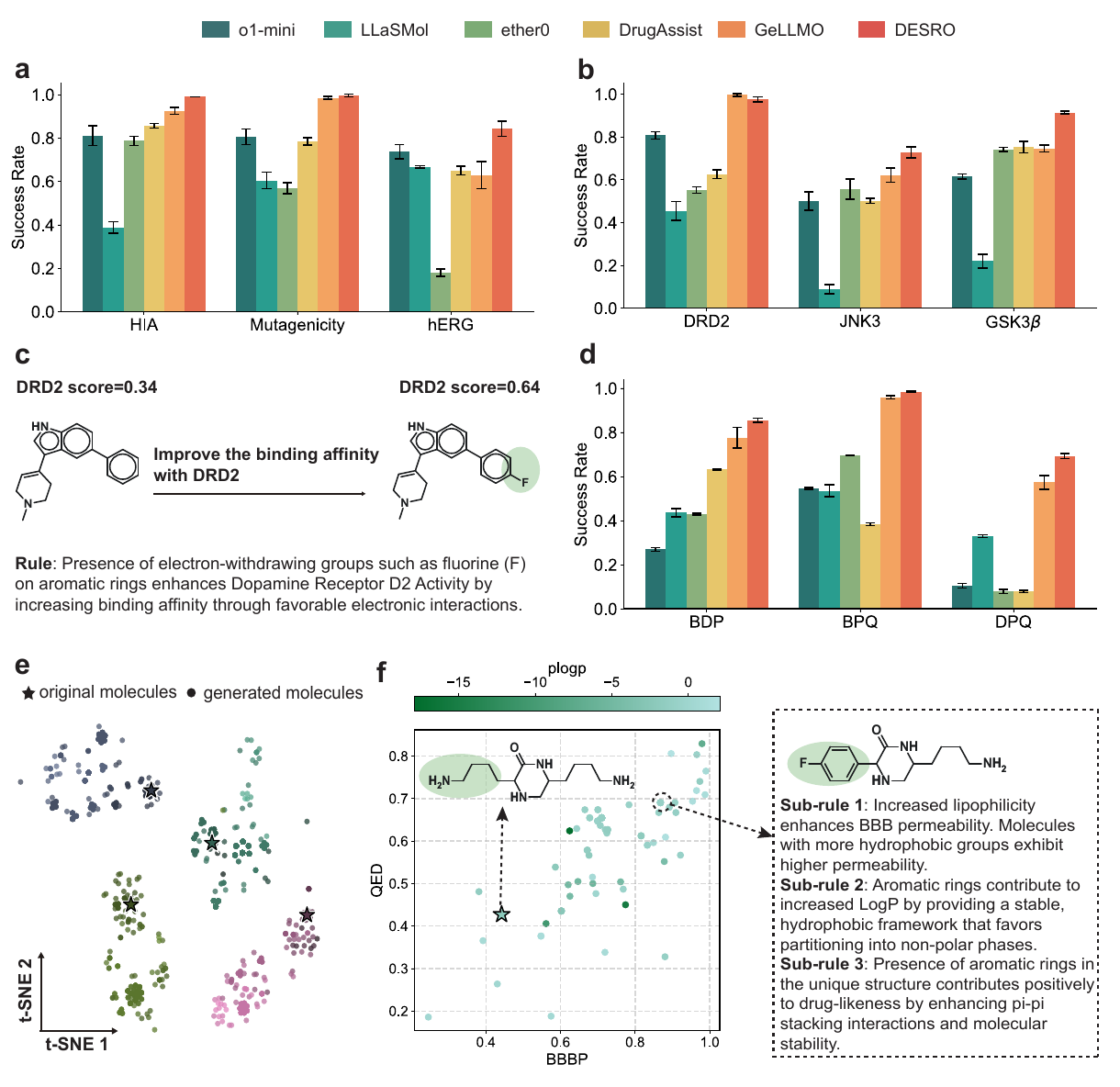} 
    \captionsetup{labelformat=empty}
    \caption{\textbf{Figure 2. Evaluation of molecule optimization on ADMET properties and bioactivity.}
\textbf{a,b, }Bar plots comparing {\ourM} and competing methods in single-property optimization tasks for ADMET properties (\textbf{a}) and bioactivity targets (\textbf{b}) in terms of Success Rate.
\textbf{c, }Case study for bioactivity optimization for the DRD2 target. The original molecule is modified based on the inferred rule regarding electron-withdrawing groups to improve binding affinity.
\textbf{d, }Bar plots comparing the success rates of {\ourM} and baselines across multi-property optimization tasks.
\textbf{e, }t-SNE plot visualizing the chemical space of original molecules (stars) and generated candidates (circles) for the BPQ (BBBP, pLogP and QED) task. Points are colored by the average score of the target properties, where lighter colors indicate a better average performance across all objectives.
\textbf{f, }Scatter plot showing the property distribution of generated molecules for a selected starting scaffold in the BPQ task in terms of QED and BBBP, colored by pLogP values. The dashed box details the sub-rules applied by the model to satisfy multiple property constraints simultaneously.}
    \label{fig:2}
\end{figure}
\noindent Optimizing ADMET properties and bioactivity is a central challenge in drug discovery, requiring models to learn complex structure-activity relationships from experimental data rather than applying simple atomic modifications. We evaluated the performance of \ourM{} on both single-property and multi-property optimization tasks encountered during training. For each input molecule, the model generated four candidates for single-property optimization and twenty candidates for multi-property optimization. We report success rate, defined as the percentage of input molecules for which at least one successful optimization was achieved. Since the properties of the generated molecules may not have been experimentally measured, we followed prior work \cite{wu2024leveraging,ye2025drugassist,dey-etal-2025-mathtt} by employing a property predictor based on machine learning to estimate their property values. We compared \ourM{} against the general-purpose reasoning model o1-mini, two chemistry LLMs, ether0 \cite{narayanan2025training} and LLaSMol \cite{yullasmol}, and two molecule optimization LLMs fine-tuned on molecule optimization pairs, DrugAssist \cite{ye2025drugassist} and GeLLMO \cite{dey-etal-2025-mathtt}.

We first benchmarked the performance of \ourM{} on three representative ADMET optimization tasks: improving human intestinal absorption (HIA), reducing mutagenicity, and reducing hERG inhibition (\textbf{Fig. \ref{fig:2}a}). \ourM{} achieves the highest success rate across all three properties, demonstrating its superior capability of molecule optimization.
\ourM{} surpasses o1-mini, the model used to generate its training data, suggesting that the reasoning induction successfully endowed our model with optimization capabilities that transcend those inherent in o1-mini. 
In general, molecule optimization LLMs outperform the general-purpose LLM, which surpasses chemical LLMs, underscoring that effective molecule optimization requires specialized training and cannot be fully acquired through pretraining on general chemical corpora alone.
While DrugAssist and GeLLMO were fine-tuned on hERG and HIA/mutagenicity respectively, \ourM{} achieves superior performance on these specific properties. Furthermore, \ourM{} outperforms Prompt-MolOpt, a specialist model employing carefully trained property embeddings, on hERG and mutagenicity, proving its efficacy as a general-purpose expert (\textbf{Supplementary Fig. \ref{fig:prompt-molopt}}). In contrast, the performance drop of DrugAssist and GeLLMO on other properties underscores the necessity for developing more generalizable molecule optimization models.

We next evaluated our model for bioactivity optimization on DRD2, JNK3, and GSK3$\beta$ (\textbf{Fig. \ref{fig:2}b}). The input instruction contained only the target name, a structure-free setting that reflects real-world scenarios where target structural information is often unavailable. Our model outperforms all baselines on JNK3 and GSK3$\beta$ and performs comparably to GeLLMO on DRD2, despite the latter being heavily fine-tuned on that specific target. A case study demonstrates that the model successfully optimizes a molecule's bioactivity on DRD2 by incorporating the electron-withdrawing atom, fluorine (\textbf{Fig. \ref{fig:2}c}). This indicates that the model effectively learns key substructures through the reasoning induction process even without access to the binding pocket structure, highlighting its broad potential for bioactivity optimization in scenarios where target structural information is unavailable.

We further evaluated our model in the more challenging and realistic setting of multi-property optimization, which better reflects real-world drug discovery objectives. We adopted four real-world multi-property optimization tasks from GeLLMO (\textbf{Fig. \ref{fig:2}d, Supplementary Fig. \ref{fig:bdq}}): (1) improving BBBP, DRD2, pLogP (BDP), (2) improving BBBP, pLogP, QED (BPQ), (3) improving DRD2, pLogP, QED (DPQ), and (4) improving BBBP, DRD2, QED (BDQ). Our model achieves state-of-the-art performance on the BDP, BPQ, and DPQ tasks, demonstrating its strong ability to generalize to multi-objective problems .
A qualitative analysis was performed to further understand this superior performance (see Methods). For four molecules in the BPQ task, we demonstrated the t-SNE \cite{maaten2008visualizing} visualization of the generated molecules. The model generated candidates that explored the local chemical space through minimal yet effective structural modifications, with a clear preference toward directions that yielded improvement in the target property. (\textbf{Fig. \ref{fig:2}e}). A large proportion of generated molecules successfully satisfied all property constraints (\textbf{Fig. \ref{fig:2}f, Supplementary Fig. \ref{fig:scatter}}). A representative case study suggests that the model implicitly learns a set of sub-rules, each targeting a specific property. By applying these modifications compositionally, the model successfully navigates the multi-objective landscape. Collectively, the strong performance of \ourM{} on both single-property and multi-property tasks demonstrates its successful training as a reasoning-based molecule optimization model, motivating us to examine its ability to generalize to out-of-domain optimization tasks.

\subsection*{Reasoning facilitates generalization to out-of-distribution scenarios}

\begin{figure}[!htp]
    \centering
    \includegraphics[width=1.0\linewidth]{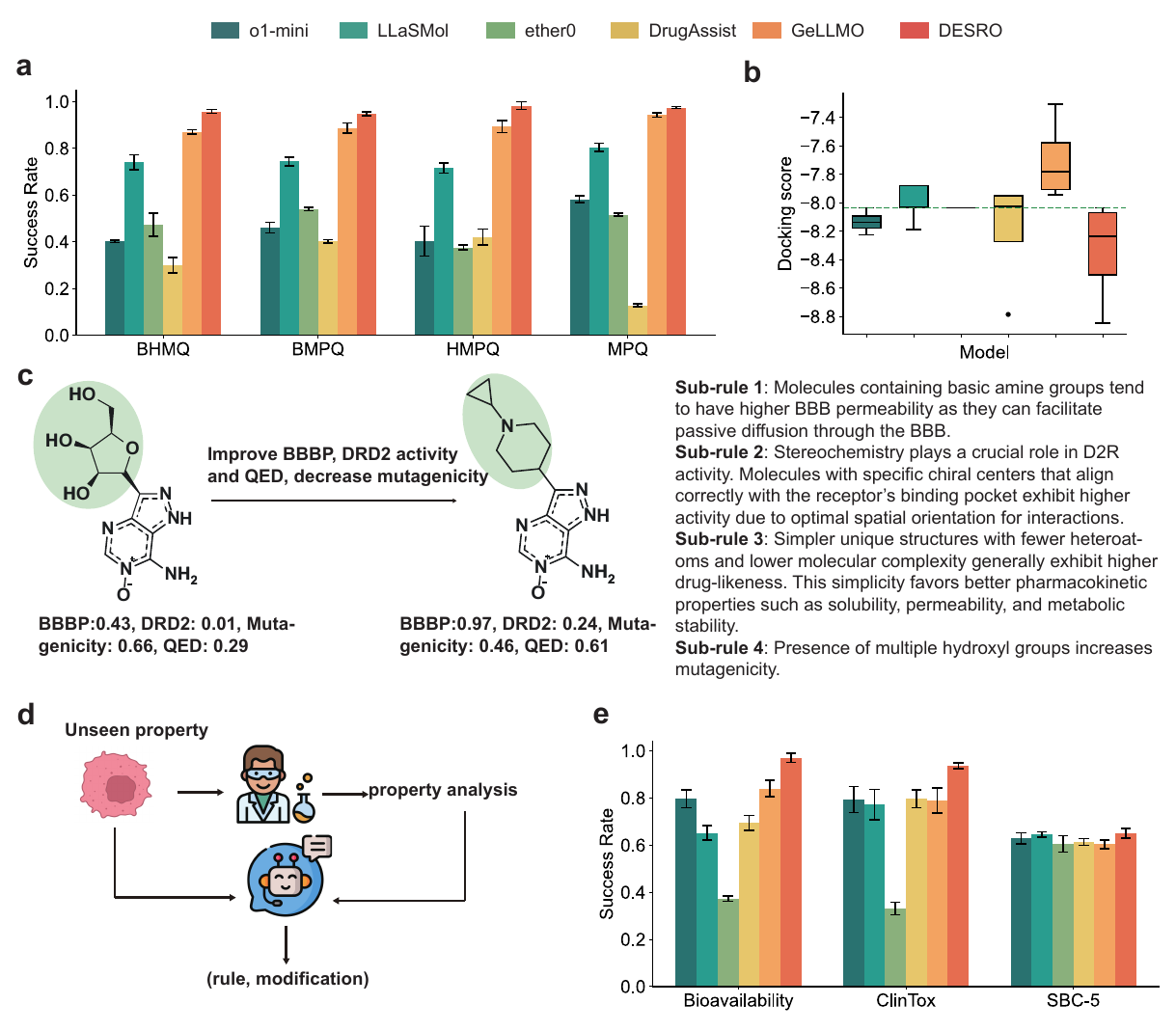} 
    \captionsetup{labelformat=empty}
    \caption{\textbf{Figure 3. Evaluation of generalization to out-of-domain (OOD) molecule optimization tasks.}
\textbf{a, }Bar plots comparing the success rates of {\ourM} and baselines on OOD multi-property optimization tasks.
\textbf{b, }Box plots showing the docking score distribution for molecules optimized for ClpP2, a protein target not encountered during training. The green dashed line indicates the docking score of the input molecule; lower scores indicate better binding affinity.
\textbf{c, }Case study of OOD multi-property optimization. The model successfully optimizes a molecule on the OOD multi-property optimization by applying learned sub-rules for single properties.
\textbf{d, }The inference workflow for optimizing entirely unseen properties. The model leverages expert-provided property analyses to generate optimization rules and modifications without prior training on those specific properties.
\textbf{e, } Bar plots comparing the success rates on three unseen properties: Oral Bioavailability, Clinical Toxicity (ClinTox), and drug sensitivity on the SBC-5 cell line. }
    \label{fig:3}
\end{figure}
\noindent As real-world drug discovery constantly involves novel molecule optimization requirements, a molecule optimization model should generalize to out-of-distribution (OOD) tasks. We assessed \ourM{} on three types of OOD tasks: OOD multi-property combinations, OOD target for bioactivity optimization and novel properties not encountered during training. 

We first evaluated our model on OOD multi-property optimization following the setting of GeLLMO (\textbf{Fig.  \ref{fig:3}a, Supplementary Fig. \ref{fig:ood_multi}}). The evaluation is conducted on five unseen combinations of ADMET properties, where the individual properties were observed during training but their specific combinations were not (see Methods). Our model significantly outperforms all baselines on four combinations while performing comparably on the remaining one, demonstrating strong generalization capabilities. Notably, it achieves the best performance on the four-property optimization tasks despite never encountering combinations of more than three properties during training, highlighting its ability to extrapolate to more complex scenarios. A case study provides insight into this behavior: the sub-rules generated by our model for the multi-property task show a strong alignment with the learned rules for single properties (\textbf{Fig. \ref{fig:3}c}). This suggests that our model has learned to solve novel problems by composing its knowledge of simpler, constituent rules. This emergent compositional reasoning is of profound practical significance for molecular discovery. Given that acquiring molecules with a full suite of desired properties is experimentally expensive and thus data is inherently scarce, our model's ability to compose knowledge from readily available single-property data provides a scalable and data-efficient pathway to navigate previously intractable multi-objective optimization challenges.

We then evaluated our model on an OOD bioactivity optimization task targeting ClpP2 in mycobacterium tuberculosis, a protein not encountered during training (\textbf{Fig. \ref{fig:3}b}). We selected a molecule with low initial bioactivity (see Methods) as a starting point and generated optimized molecules. The results showed that all of our generated molecules achieved improved docking scores, while baseline methods exhibited minimal success. Since the target was entirely novel to the model, \ourM{} could not utilize target-specific binding information. However, its successful optimization of the majority of molecules indicates that it has learned generalizable principles of bioactivity applicable across diverse protein targets.

Finally, we evaluated our model in a more challenging setting where the property to be optimized was entirely unseen during training. To simulate the real-world drug discovery scenario, the model is provided with the property name and an expert-curated natural language description of the property as input, and outputs the optimization rules and modification accordingly (\textbf{Fig. \ref{fig:3}d}). We tested this on two critical ADMET properties, oral bioavailability and toxicity in clinical trials, as well as drug sensitivity on the SBC-5 cell line, a standard model in preclinical cancer research (\textbf{Fig. \ref{fig:3}e}). Collectively, these properties play a pivotal role in advancing compounds from lead identification to preclinical development. Our model consistently outperforms the competing methods, demonstrating its ability to generalize to novel optimization objectives defined purely by textual descriptions. In contrast, other molecule optimization LLMs showed limited performance gains compared with general-purpose LLMs, highlighting their limitation to adapt to diverse optimization settings. This underscores the potential of \ourM{} for real-world discovery in flexible and low-resource drug discovery settings. 

\subsection*{Navigating real-world chemical space for lead optimization}

\begin{figure}[!htp]
    \centering
    \includegraphics[width=1.0\linewidth]{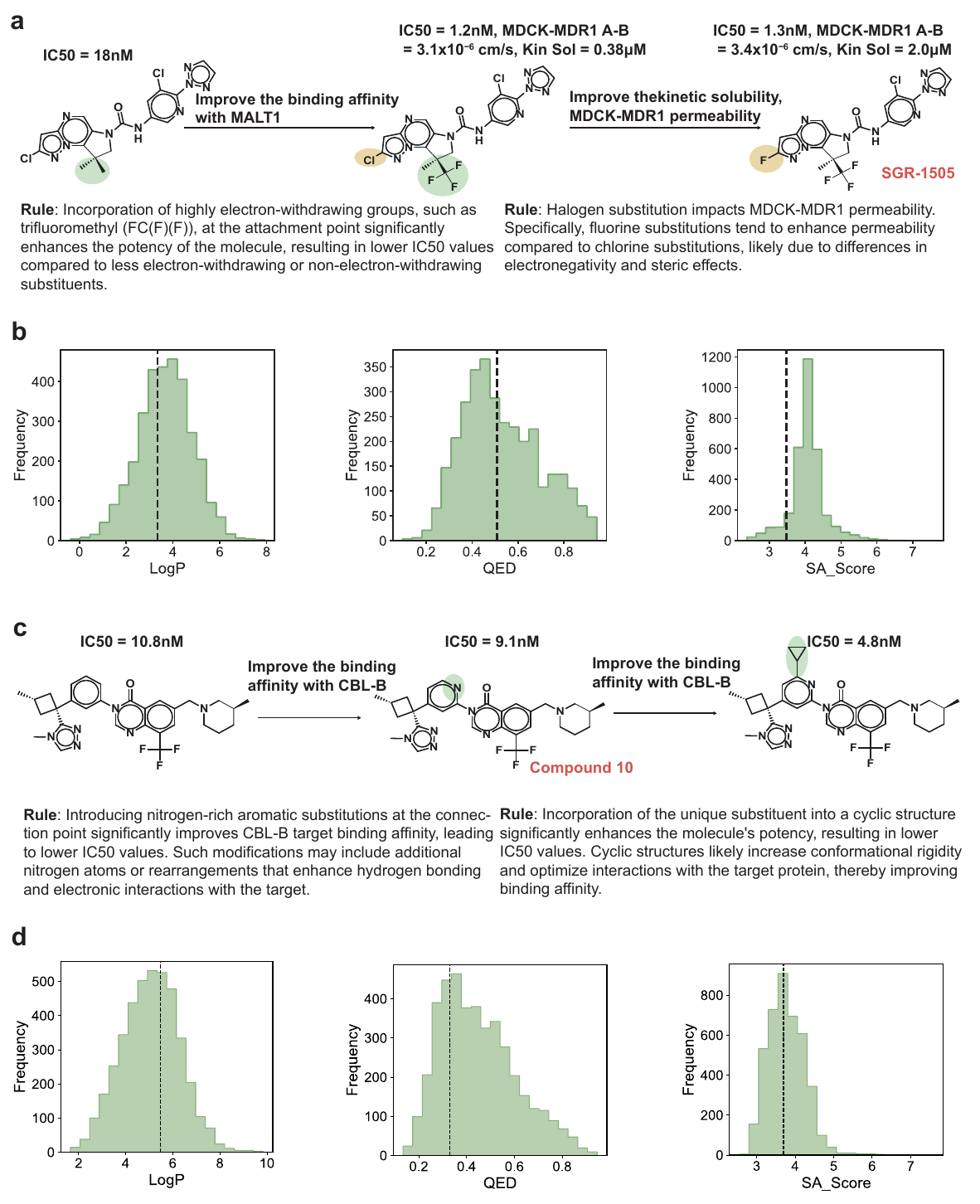} 
    \captionsetup{labelformat=empty}
    \caption{\textbf{Figure 4. Evaluation of lead optimization in real-world drug discovery scenarios.}
\textbf{a, }Case study on the optimization of MALT1 inhibitors. {\ourM} successfully guides optimization from an initial lead to the clinical candidate SGR-1505. 
\textbf{b,d, }Histograms illustrating the physicochemical property distributions of the molecular libraries generated by {\ourM} for the MALT1 (\textbf{b}) and CBL-B (\textbf{d}) optimization tasks. The distributions cover LogP, QED, and Synthetic Accessibility (SA) Score. The vertical dashed lines represent the property values of the input molecules.
\textbf{c, } Case study on the optimization of CBL-B inhibitors. The model navigates the chemical space from a hit compound to the lead molecule "Compound 10" and subsequently identifies a structurally distinct analog with superior potency.}
    \label{fig:4}
\end{figure}
\noindent The promising results of \ourM{} in out-of-domain molecule optimization tasks further motivated us to apply it in real-world drug discovery scenarios. We selected two lead optimization tasks \cite{nie2025accelerated,meng2025discovery} published after October 2025, while all training data were curated from datasets released no later than October 2025, ensuring that these case studies were entirely temporally held out and unseen during training.

Targeting MALT1 has emerged as a promising therapeutic strategy for treating mature B-cell malignancies. In the optimization of allosteric inhibitors, \ourM{} autonomously guided the chemical evolution from an initial lead to the clinical development candidate SGR-1505 with IC$_{50}$ value of 1.3 nM \cite{nie2025accelerated} (\textbf{Fig. \ref{fig:4}a}). First, we employed \ourM{} to improve the binding affinity for MALT1 of the initial lead. The model exhibited expert-level chemical intelligence by successfully driving a 15-fold increase in binding affinity by introducing a highly electron-withdrawing substituent, supported by an explicit reasoning process. Then the model navigated the late-stage optimization to refine the kinetic solubility and permeability while maintaining the binding affinity. Furthermore, the generated molecular library exhibits physicochemical property distributions, including LogP, QED, Synthetic Accessibility (SA) scores and molecular weight, that align closely with those of the initial lead (\textbf{Fig. \ref{fig:4}b}, \textbf{Supplementary Fig. \ref{fig:mw}}). This demonstrated that \ourM{} can achieve substantial potency improvements while preserving key drug-like properties and synthetic feasibility within the lead series.

Extending our evaluation to the field of cancer immunotherapy, we applied \ourM{} to the optimization of CBL-B inhibitors (\textbf{Fig. \ref{fig:4}c}), where pharmacological modulation of this E3 ubiquitin ligase represents a promising strategy for enhancing T-cell–mediated antitumor immunity. \ourM{} autonomously navigated the chemical space from a hit compound to the lead molecule with IC$_{50}$ value of 9.1 nM \cite{meng2025discovery}  by proposing nitrogen-rich aromatic substitution. The physicochemical distributions of the generated library also align closely with the reference hit (\textbf{Fig. \ref{fig:4}d}, \textbf{Supplementary Fig. \ref{fig:mw}}), highlighting the potential of \ourM{} to generate novel and chemically rational molecules. By further prioritizing the optimization of binding affinity, \ourM{} successfully identified a structurally distinct analog exhibiting superior biochemical potency. Although this compound was eventually de-prioritized due to suboptimal metabolic stability, its discovery underscores the capacity of \ourM{} to exhaustively explore the chemical space and identify candidates with maximal target engagement. Collectively, these findings highlight the practical utility of \ourM{} in streamlining real-world drug design cycles, effectively reducing the substantial manual and computational costs associated with identifying high-quality clinical candidates.

\begin{figure}[h]
    \centering
    \includegraphics[width=1.0\linewidth]{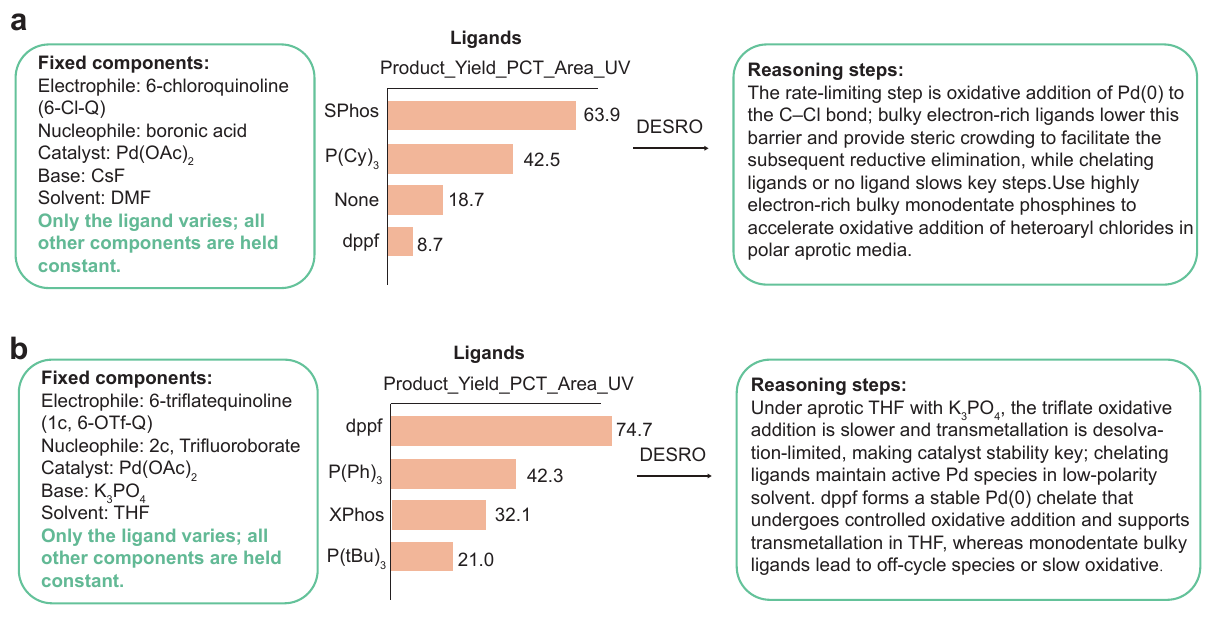} 
    \captionsetup{labelformat=empty}
    \caption{\textbf{Figure 5. Extension to reaction ligand selection.} \textbf{a,} Ligand selection for a heteroaryl chloride (6-chloroquinoline), where \ourF{} infers (based on controlled comparisons within the dataset) that oxidative addition is a plausible barrier and prioritizes a bulky, electron-rich monodentate ligand (SPhos). \textbf{b,} Ligand selection for a heteroaryl triflate (6-triflatequinoline) in THF, where \ourF{} emphasizes catalyst speciation/stability considerations and selects a chelating bidentate ligand (dppf). These examples illustrate condition-dependent reasoning for ligand choice.}
    \label{fig:5}
\end{figure}
\subsection*{Exploratory extension to reaction ligand selection}
\noindent To demonstrate that \ourF{} generalizes beyond molecular structure optimization, we applied the framework to ligand selection in transition-metal catalysis, a critical component of organic synthesis where identifying an optimal ligand requires reasoning about coupled effects of the substrate, solvent, base, and catalyst speciation/stability. Using the high-throughput Suzuki–Miyaura cross-coupling dataset reported by Perera et al. \cite{perera2018platform}, we constructed controlled comparison groups in which the substrate and reaction conditions are held fixed while only the ligand varies, enabling data-driven induction of ligand–reactivity reasoning. For a heteroaryl chloride substrate (\textbf{Fig. \ref{fig:5}a}), \ourM{} generates a rationale consistent with oxidative addition being a major kinetic challenge in this regime and prioritizes a bulky, electron-rich monodentate ligand (SPhos). In contrast, for a heteroaryl triflate in THF (\textbf{Fig. \ref{fig:5}b}), \ourM{} produces a context-aware rationale that emphasizes catalyst robustness/speciation and selects a chelating bidentate ligand (dppf), rather than simply maximizing ligand electron donation. These cases illustrate that our framework can autonomously decipher the latent reasoning in chemical reactivity, suggesting broad potential for interpretable experimental design in diverse scientific domains.
\section*{Discussion}
\noindent We have presented a framework, \ourF{}, that deciphers intermediate scientific reasoning from outcome data via groupwise comparison. It enables the scalable reconstruction of an interpretable reasoning process for molecule optimization from large-scale molecular property datasets. Building on the derived data, we have presented a reasoning model for molecule optimization. The model performs molecule optimization through a structured reasoning process, integrating property analysis, rule induction, and molecular modification. We have demonstrated its strong performance across a range of tasks, including single-property and multi-property optimization, as well as generalization to out-of-distribution settings such as unseen properties and targets. We further validated its real-world utility by retrospective case studies of inhibitor design for MALT1 and CBL-B. Additionally, we show that the paradigm can be extended to another scientific design problem, reaction ligand selection. \ourM{} offers a scalable and generalizable solution for building scientific reasoning models, and holds strong potential to accelerate data-driven scientific discovery.

\ourF{} complements existing scientific reasoning models \cite{wang2025scireasoner,narayanan2025training,edwards2025mclm,fallahpour2025bioreason,wang2025chem,zhao2025chemdfm} by providing a scalable source of supervised data. It captures reasoning patterns that are often absent from the pretraining corpora of general-purpose LLMs, by reconstructing stepwise rationales directly from experimental outcomes. This LLM-assisted reconstruction enables rapid adaptation to new scientific domains. Notably, on molecule optimization, fine-tuning only a LLaMA3-8B backbone yields performance that can even surpass strong general reasoning models such as o1-mini, highlighting the potential of outcome-derived reasoning supervision to strengthen scientific reasoning. Additionally, it effectively extends the AI-driven scientific discovery paradigm \cite{boiko2023autonomous,penades2025ai,zheng2025large} to the autonomous recovery of latent reasoning trajectories for every individual entry in large-scale datasets. While the current study primarily validates \ourF{} through quantitative experiments on molecule optimization, the successful application to catalysis selection demonstrates the potential of this framework to generalize to broader scientific tasks, representing a promising direction for future work.

Machine learning and deep learning methods have been explored for molecule optimization \cite{graff2021accelerating,korovina2020chembo,yoshikawa2018population,nigamaugmenting,nigam2021beyond, jensen2019graph,gaoamortized,gao2024generative,jin2018junction,maus2022local,de2018molgan,xiemars,lee2023exploring,lee2024molecule,chen2025uncertainty,zhu2023sample,liu2024graph,lee2025genmol,he2021molecular,fu2021mimosa,brown2019guacamol,olivecrona2017molecular,zhou2019optimization,you2018graph,jin2020multi,yang2021hit,horwood2020molecular,fu2022reinforced,dodds2024sample,bou2024acegen,guo2025directly,yoshizawa2025data,shen2021deep,fudifferentiable,niuinversiongnn}. Compared to these methods, our model does not require training task-specific generators or predictors. Instead, it employs an LLM to jointly optimize multiple properties and generalize to arbitrary new objectives via textual prompts. Previous LLM-based molecule optimization approaches can be classified into two categories, including LLMs finetuned on molecule optimization pairs \cite{dey-etal-2025-mathtt,zhu2025conditional,dey2025large,wu2024leveraging,ye2025drugassist,edwards2025mclm,liu2023multi,narayanan2025training,loeffler2024reinvent,wang2025polo,xia2025nature} and general-purpose LLMs with prompt engineering or agentic workflow \cite{wangefficient,nguyenlico,xia2024evolutionary,liu2023chatgpt,kim2025mt,gao2025pharmagents,yu2025collaborative,zhu2025coder}. Compared to finetuned LLMs, our model explicitly captures the step-by-step reasoning process underlying molecule optimization, offering stronger interpretability and generalization. Compared to general-purpose LLMs, our model learns from optimization rationales derived by \ourF{}, establishing a new paradigm for applying LLMs to molecule optimization.

While the results of \ourM{} are promising, it still has several limitations. First, the data construction process is currently based on o1-mini; consequently, this framework can be readily extended to incorporate more advanced reasoning models for training data construction. Second, the current formulation of bioactivity optimization does not include target structure information, which will be integrated in future work to enable more precise optimization. 
Finally, reinforcement learning (RL) is not yet employed in \ourM{}. Our current focus is on establishing foundational reasoning abilities of LLMs for molecule optimization. Incorporating RL methods such as GRPO \cite{guo2025deepseek} for self-improvement of the reasoning process will be a key direction in future work.

\section*{Methods}
\subsection*{Framework of \ourF{}}
\noindent We assume access to outcome-only records $\mathcal{D}=\{(x_i, y_i)\}_{i=1}^{N}$, where $x_i$ denotes a design instance (e.g., a molecule) and $y_i$ is its measured outcome under a target criterion (e.g., molecular property value). A grouping function $g(\cdot)$ partitions instances into groups that share a common context or core attributes, yielding $\mathcal{G}_k=\{i \mid g(x_i)=k\}$. For each group $k$, we summarize all within-group observations into a structured set
\begin{equation}
S_k=\Big\{\big(x_i, y_i\big)\ \big|\ i\in \mathcal{G}_k\Big\},
\end{equation}
augmented with derived descriptors such as a shared core representation and the corresponding group-specific edits or differences. We model the intermediate reasoning trace $r_k$ as a latent variable that explains the observed pattern of outcome variation within the group. Rather than relying on human-annotated rationales, we infer $r_k$ directly from $S_k$. Concretely, we prompt a large language model (LLM) with $S_k$ to produce a reasoning trace that best explains the shared patterns and key differences in the group.


\subsection*{Problem setting of molecule optimization}
\noindent
The goal of molecule optimization is to generate a new molecule \( m' \) from a given input molecule \( m  \) and an optimization instruction \( I  \), such that the optimized molecule improves upon the original with respect to the specified property. Formally, this can be modeled as learning a function \( f \), where \( m' = f(m, I) \) such that \( v(m') > v(m) \), with $v(m)$ denoting the property value of $m$. 

\subsection*{Reasoning model for molecule optimization}

\noindent We instantiates \( f \) as a reasoning-based large language model that performs molecule optimization through a three-step process. First, it conducts a property analysis by generating a textual summary \( a \) describing the property definition and its structural or physicochemical determinants. Second, it induces an optimization rule \( r \), suggesting an optimization strategy based on the property analysis and the input molecule structure. Third, based on the induced rule $r$, the model decomposes the original molecule $m$ into a preserved fragment $s_{\text{keep}}$ and its complementary substructure $s_{\text{rep}} = m \setminus s_{\text{keep}}$. The model then proposes a new fragment $s_{\text{new}}$ to replace $s_{\text{rep}}$, yielding the optimized molecule $\hat{m} = \text{Combine}(s_{\text{keep}}, s_{\text{new}})$. This structured pipeline enables our model to generate not only optimized molecules but also reasoning paths aligned with the optimization goal.

\subsection*{Molecular property dataset curation}
\noindent We constructed a large-scale molecular property dataset covering both bioactivity and ADMET endpoints. For bioactivity, we obtained assays from PubChem BioAssay\footnote{\url{https://ftp.ncbi.nlm.nih.gov/pubchem/Bioassay/}}
. In total, this yielded 1.4M molecule–bioactivity pairs. 
For ADMET properties, we combined the training splits from Prompt-MolOpt \cite{wu2024leveraging} and GeLLMO, including: blood–brain barrier permeability (BBBP), water solubility, hERG inhibition, human intestinal absorption (HIA), mutagenicity, penalized octanol–water partition coefficient (pLogP), and quantitative estimate of drug-likeness (QED), resulting in 589k molecule–property pairs. We additionally incorporated 154k molecules annotated with hydrogen-bond acceptors (HBA) and hydrogen-bond donors (HBD) from DrugAssist \cite{ye2025drugassist} to strengthen the model’s grasp of basic physicochemical features. We filtered out all the molecules in the evaluation set to avoid data leakage. The final dataset is organized by assay; for synthetic data, each data source is treated as an assay. Each assay is associated with a metadata record, including a textual assay description and a directionality flag (i.e., \textit{higher-is-better} or \textit{lower-is-better}) indicating whether larger or smaller values correspond to improved property outcomes. For assays from PubChem, the assay descriptions were downloaded from \url{https://ftp.ncbi.nlm.nih.gov/pubchem/Bioassay/Extras/bioassays.tsv.gz}. For other properties, the descriptions were manually curated.

\subsection*{Common fragment mining}
\noindent We follow the connection-aware motif mining in MiCaM to derive a compact set of recurring fragments directly from data~\cite{gengnovo}. In brief, MiCaM learns a short sequence of data-driven merge operations by repeatedly picking the most frequent adjacent-fragment merge across the corpus and applying it to all molecules. Applying the learned sequence decomposes each molecule into a small number of connection-aware motifs, i.e., fragments that carry explicit attachment sites so their typical bonding context is preserved. This avoids hand-crafted rules and yields a vocabulary that reflects how fragments actually connect in real molecules.

For each assay, we run the miner on our curated training molecules and collect the most frequent connection-aware fragments as the motif vocabulary used by our generator and analysis pipeline. As illustrated by the PubChem example for the angiotensin II receptor (Fig.~\ref{fig:1}b), we take compounds with assay measurements and mine the high-frequency fragments that appear across them; these become reusable building blocks with well-defined attachment points (denoted as '*' in the figure). Finally, we obtain a set of frequent substructures \( \mathcal{C} = \{c_k\}_{k=1}^K \) for each assay.

\subsection*{Deciphering reasoning steps}
\noindent
\textbf{Single property optimization.} 
For each assay, we organize the training data into a set of groups $\mathcal{G} = \{G_k\}_{k=1}^K$ based on the frequent substructures $\mathcal{C} = \{c_k\}_{k=1}^K$.
Each group $G_k$ contains molecules that share the same core fragment $c_k$. Formally, $G_k = \{(m, v(m)) \mid c_k \subset m\}$, where any molecule $m \in G_k$ can be decomposed as $m = \text{Combine}(c_k, s)$, with $s$ denoting the unique substructure, and $v(m)$ representing the property value of $m$.
We provide the entire collection of groups $\mathcal{G}$ to the o1-mini model \cite{jaech2024openai} to perform rationale induction. The model inputs include the shared cores, the unique substituents, and the property values across all groups, along with the assay metadata (see \textbf{Supplementary Fig. \ref{fig:prompt}} for the prompt details).

The reasoning process proceeds in two stages. 
First, o1-mini generates a global property analysis $a$, summarizing the structural and physicochemical determinants of the property.
Second, conditioned on $a$ and the aggregated groups $\mathcal{G}$, the model induces a set of optimization rules $\mathcal{R} = \{r_j\}_{j=1}^J$. 
Crucially, for each rule $r_j$, the model identifies a set of supporting molecule pairs $P_j$ that exemplify the rule. 
A pair $(m, m')$ is valid if and only if both molecules belong to the same group (i.e., they share a common fragment $c_k$) and exhibit a property improvement. 
Although each pair is from a specific group, a rule $r_j$ aggregates evidence across different groups. The supporting set is defined as:
\[
P_j = \left\{ (m, m') \ \middle|\ 
\begin{aligned}
& \exists k \in \{1, \dots, K\} \text{ s.t. } m, m' \in G_k, \\
& m = \text{Combine}(c_k, s), \ m' = \text{Combine}(c_k, s'), \\
& v(m') > v(m), \text{ and the pair aligns with } r_j
\end{aligned}
\right\}.
\]

In this way, each pair in $P_j$ demonstrates the specific structural modification (changing $s$ to $s'$ while keeping $c_k$) mandated by rule $r_j$.

For a given molecule optimization pair $(m, m')$, the complete reasoning process includes: (1) an instruction in natural language, such as \textit{``improve the \{property\} of $m$''}; (2) a property analysis $a$; (3) the applicable optimization rules $\mathcal{R}_{(m, m')} \subseteq \mathcal{R}$ (noting that a single molecule pair may support multiple distinct rules); (4) a modification plan expressed as \textit{``we should keep $c_k$ and modify $s$''}; and (5) the resulting optimized fragment $s'$. 

\textbf{Multi-property optimization.} 
We construct reasoning trajectories for multi-objective scenarios by identifying molecule pairs $(m, m')$ that simultaneously satisfy the optimization criteria for a set of target properties $\mathcal{T} = \{p_1, \ldots, p_L\}$.
First, let $\mathbb{P}^{(\ell)}$ denote the set of all valid optimization pairs identified for property $p_\ell$.
We define the multi-property dataset as the intersection of these sets:
\[
P^{\text{multi}} = \bigcap_{\ell=1}^L \mathbb{P}^{(\ell)}.
\]
For each pair $(m, m') \in P^{\text{multi}}$, the molecule $m$ and $m'$ may have been decomposed differently for each property. Specifically, for property $p_\ell$, the pair belongs to a group whose common fragment is $c^{(\ell)}$. To derive a consistent structural transformation for the multi-property task, we take the intersection of the common structures associated with each property:
\[
c^{\text{shared}} = \bigcap_{\ell=1}^L c^{(\ell)}.
\]
Accordingly, the modifiable substituents are redefined relative to this shared scaffold: $s = m \setminus c^{\text{shared}}$ and $s' = m' \setminus c^{\text{shared}}$.
We then aggregate the textual rationales, including the global analyses $\{a^{(\ell)}\}_{\ell=1}^L$ and the specific induced rules $\{r^{(\ell)}\}_{\ell=1}^L$, into a unified rationale.
The resulting trajectory teaches the model to recognize that keeping $c^{\text{shared}}$ and modifying $s$ to $s'$ simultaneously satisfies the diverse optimization rules encoded in $\{r^{(\ell)}\}_{\ell=1}^L$.
We consider $L \leq 3$ in the training set.

\subsection*{Molecule pair selection}
\noindent
To ensure that optimization pairs reflect significant improvements, we apply a filtering criterion based on property value differences. Specifically, for each molecule pair \((m, m')\), we retain it in the training set only if the difference in property value exceeds a predefined threshold. For single-property optimization, a pair is included only if:
\[
|v(m')- v(m)| > \delta_p,
\]
where \( \delta_p \) is a property-specific threshold (\textbf{Supplementary Table \ref{tab:delta}}). 

For multi-property optimization involving a set of target properties \( \{p_1, p_2, \ldots, p_L\} \), we apply the thresholding criterion independently to each property. A molecule pair \((m, m')\) is retained only if it satisfies the threshold condition for all properties simultaneously:
\[
|v^{(\ell)}(m') - v^{(\ell)}(m)| > \delta_{p_\ell} \quad \text{for all } \ell = 1, \ldots, L,
\]
where \( v^{(\ell)}(m) \) and \( v^{(\ell)(m')} \) denote the values of property \( p_\ell \) for \( m \) and \( m' \), respectively, and \( \delta_{p_\ell} \) is the corresponding threshold. This ensures that retained pairs demonstrate consistent and meaningful improvements across all targeted objectives, rather than trading off one property against another. 
\subsection*{Training data construction}
\noindent
Each training instance for property $p$ is constructed from an optimization pair \((m, m')\) using a structured prompt. The template is defined as:

\begin{tcolorbox}[colback=gray!5!white, colframe=gray!50!black, title=Prompt Template]
\texttt{Instruction\_Template}($p$, $m$) (see \textbf{Supplementary Tables \ref{tab:prompt-bioactivity}-\ref{tab:prompt-admet}} for the full template list) \\
\texttt{<property\_analysis>} \( a \) \texttt{</property\_analysis>} \\
\texttt{<rule>} \( r \) \texttt{</rule>} \dots \quad (repeats for each \( r \in \mathcal{R}_{(m, m')} \)) \\
\texttt{<modification>} we should keep \( c_k \) and modify \( s \) \texttt{</modification>} \\
\texttt{Optimized fragment:} \( s' \)
\end{tcolorbox}

Based on this format, we define three task settings, each corresponding to a different supervised objective:

\textbf{Task 1 (Modification Generation)} includes two subtypes. 
In \textbf{Task 1.1 (Constrained Modification)}, the input is 
\[
x = (\text{instruction}, a, \mathcal{R}_{(m, m')}, \texttt{<modification>} \text{keep } c_k \text{ and modify } s \texttt{</modification>}),
\]
and the output is the optimized fragment:
\[
y = s'.
\]

In \textbf{Task 1.2 (Unconstrained Modification)}, the model is given only the instruction, property analysis, and optimization rules:
\[
x = (\text{instruction}, a, \mathcal{R}_{(m, m')}),
\]
and must generate both the modification plan and the optimized fragment:
\[
y = (\texttt{<modification>} \text{keep } c_k \text{ and modify } s \texttt{</modification>},\ s').
\]

In \textbf{Task 2 (Property-Guided Reasoning)}, the inputs consists only of the instruction and property analysis:
\[
x = (\text{instruction}, a),
\]
and the model is trained to generate both the optimization rules and the modification result:
\[
y = (\mathcal{R}_{(m, m')},\ \texttt{<modification>} \text{keep } c_k \text{ and modify } s \texttt{</modification>},\ s').
\]

\textbf{Task 3 (Full Reasoning)} is the most open-ended setting, where the model is provided only with the instruction:
\[
x = \text{instruction},
\]
and must predict the full reasoning trajectory, including the property analysis, rules, modification, and optimized fragment:
\[
y = (a,\ \mathcal{R}_{(m, m')},\ \texttt{<modification>} \text{keep } c_k \text{ and modify } s \texttt{</modification>},\ s').
\]

For each molecule pair, we generate one training sample for each of the defined tasks. To mitigate overfitting, we downsample data for Task 1 and Task 2. Data from Task 1.1, Task 1.2, Task 2, Task 3 are mixed according to a fixed ratio of 1:2:2:5.

\subsection*{Model training}
\noindent
We fine-tune a LLaMA3-8B model on the constructed reasoning-based molecule optimization dataset using the supervised fine-tuning (SFT) paradigm. Given an input–output pair \( (x, y) \), where \( x \) denotes the prompt and \( y \) is the corresponding expected output, the model is trained to minimize the standard language modeling loss:

\[
\mathcal{L}_{\text{SFT}} = - \sum_{t=1}^{T} \log P_\theta(y_t \mid y_{<t}, x),
\]

where \( y_t \) denotes the \( t \)-th token in the output sequence, and \( P_\theta \) is the probability distribution defined by the model with parameters \( \theta \).
We train the model for 2 epochs with a learning rate of \( 2 \times 10^{-5} \), using the AdamW \cite{loshchilovdecoupled} optimizer.

\subsection*{Evaluation tasks}
\noindent We established a comprehensive benchmark to evaluate the optimization capabilities of \ourM{} across both in-distribution properties (i.e., properties encountered during training) and out-of-distribution properties.

\begin{itemize}
    \item \textbf{In-domain property optimization:}
    \begin{itemize}
        \item \textbf{Bioactivity optimization:} Optimization was benchmarked against three well-established therapeutic targets: DRD2 (Dopamine Receptor D2), JNK3 (c-Jun N-terminal kinase 3), and GSK3$\beta$ (Glycogen synthase kinase 3 beta). The test set was sourced from ChemCoTBench~\cite{li2025chemicalqaevaluatingllms}. Each target contains 100 molecules to be optimized.
        
        \item \textbf{ADMET optimization:} Evaluation was performed on three foundational ADMET properties: HIA (Human Intestinal Absorption), Mutagenicity, and hERG inhibition. For HIA and Mutagenicity, we sampled 100 molecules from the evaluation set of GeLLMO. For hERG, we sampled 100 molecules from the evaluation set of DrugAssist.
        
        \item \textbf{Multi-property optimization:} We employed the evaluation tasks from GeLLMO~\cite{dey-etal-2025-mathtt}, involving 4 optimization tasks: (1) BDP (increase BBBP, DRD2, pLogP), (2) BPQ (increase BBBP, pLogP, QED), (3) DPQ (increase DRD2, pLogP, QED), and (4) BDQ (increase BBBP, DRD2, QED). The in-domain task BDPQ was not included because the training task of our model only contains multi-property combinations with fewer than four properties.
    \end{itemize}

    \item \textbf{Out-of-domain property optimization:}
    This category evaluated the zero-shot generalization capabilities of \ourM{} on tasks and properties not included in its training data.
    \begin{itemize}
        \item \textbf{Unseen multi-property combinations:} We employed the out-of-distribution (OOD) tasks from GeLLMO~\cite{dey-etal-2025-mathtt}, which challenge the model with novel property combinations: (1) MPQ (increase pLogP, QED and decrease Mutagenicity), (2) BDMQ (increase BBBP, DRD2, QED and decrease Mutagenicity), (3) BHMQ (increase BBBP, HIA, QED and decrease Mutagenicity), (4) BMPQ (increase BBBP, QED, pLogP and decrease Mutagenicity), and (5) HMPQ (increase HIA, pLogP, QED and decrease Mutagenicity).
        
        \item \textbf{Unseen ADMET properties and drug sensitivity:} We assessed generalization to novel ADMET properties: Oral Bioavailability, Clinical Toxicity (ClinTox), and drug sensitivity on the SBC-5 cell line. For Bioavailability and ClinTox, we sampled 100 molecules from the TDC benchmark~\cite{Huang2021tdc} with low Oral Bioavailability and high Clinical Toxicity respectively. For SBC-5, we selected molecules from the GDSC dataset~\cite{10.1093/nar/gks1111} with pIC50 $\leq 5$. For this task, we provided the model with a prompt containing instruction and expert-curated property analysis to guide the optimization process on unseen properties. 
    \end{itemize}
\end{itemize}
\subsection*{Evaluation pipeline}
\noindent To ensure a standardized and reproducible comparison, all models were evaluated using a consistent pipeline, including the evaluation metric and external property predictors.

\textbf{Metrics and Success Criteria.} The primary metric for all tasks is the \textbf{Success Rate (SR)}. This is defined as the percentage of source molecules for which a model generates at least one candidate molecule that meets the specified improvement criteria.
For each source molecule, 4 candidate molecules were generated for single-property tasks, and 20 candidates were generated for multi-property tasks. The improvement criteria are:
\begin{itemize}
    \item \textbf{Single-property:} A generated molecule must have a better property score than the original source molecule. ``Better" is defined as an \textbf{increase} in the score for DRD2, JNK3, GSK3$\beta$, HIA, Bioavailability, and pIC50, or a \textbf{decrease} in the score for hERG, Mutagenicity, and ClinTox.
    \item \textbf{Multi-property:} The SR metric is stricter, requiring a generated molecule to show simultaneous improvement across \textit{all} specified properties.
\end{itemize}

\textbf{Property predictors.} We employed established property prediction models to estimate the properties of generated and reference compounds (\textbf{Table~\ref{tab:predictors}}).

\begin{table}[h!]
\centering
\caption{Summary of property predictors and their reported reliability.}
\label{tab:predictors}
\small
\begin{tabular}{@{}l l l l@{}}
\toprule
\textbf{Category} & \textbf{Property} & \textbf{Predictor / Scorer} & \textbf{Reliability} \\
\midrule
\multicolumn{4}{@{}l}{\textit{\textbf{In-Domain Properties}}} \\
\multirow{3}{*}{Bioactivity} & DRD2 & \multirow{3}{*}{\texttt{tdc.Oracle}~\cite{Huang2021tdc}} & AUROC=$0.99$ \\
 & JNK3 & & AUC = 0.983 \\
 & GSK3$\beta$ & & AUC = 0.984 \\
\addlinespace 
\multirow{4}{*}{ADMET} & BBBP & \multirow{4}{*}{ADMET-AI Platform~\cite{Swanson2024-qb}} & AUROC = $0.869 \pm 0.027$ \\
 & HIA & & AUROC=$0.981 \pm 0.002$ \\
 & Mutagenicity & & AUROC=$0.850 \pm 0.004$ \\
 & hERG & & AUROC=$0.840 \pm 0.007$ \\
\addlinespace
\multirow{2}{*}{Physicochemical} & QED & \multirow{2}{*}{RDKit-based functions} & N/A (Deterministic) \\
 & pLogP & & N/A (Deterministic) \\
\midrule
\multicolumn{4}{@{}l}{\textit{\textbf{Out-of-Domain Properties}}} \\
\multirow{2}{*}{ADMET} & Bioavailability & \multirow{2}{*}{ADMET-AI Platform~\cite{Swanson2024-qb}} & AUROC=$0.667
 \pm 0.068$ \\
 & ClinTox & & AUROC=$0.874\pm0.066$ \\
\addlinespace
Cell Response & SBC-5 pIC50 & ActFound~\cite{feng2024bioactivity} & $R^2=0.757$ \\
\bottomrule
\end{tabular}
\end{table}
\subsection*{Comparative methods}
\noindent We benchmarked \ourM{} against a suite of state-of-the-art large language models (LLMs) for molecule optimization. These baselines represent distinct architectures and training paradigms:

\begin{itemize}
    \item \textbf{o1-mini:} A general-purpose reasoning model from OpenAI optimized for STEM fields, demonstrating advanced chemical reasoning capabilities.
    
    \item \textbf{LLaSmol:} A Mistral-7B model fine-tuned on \textit{SMolInstruct}, a large-scale (3M+ samples) instruction dataset. \textit{SMolInstruct} is specifically designed to teach LLMs to follow diverse, chemically relevant instructions, including molecule optimization~\cite{yullasmol}.
    
    \item \textbf{ether0:} A 24-billion parameter chemistry reasoning model. ether0 was trained using reinforcement learning on a proprietary dataset of over 640,000 experimental chemistry problems, endowing it with strong reasoning capabilities for structure-property relationships~\cite{narayanan2025training}.
    
    \item \textbf{DrugAssist:} An interactive, Llama2-7B-Chat-based model fine-tuned on the \textit{MolOpt-Instructions} dataset. It is specifically designed to perform molecule optimization in a conversational, human-in-the-loop dialogue format~\cite{ye2025drugassist}.
    
    \item \textbf{GeLLMO:} A series of models specifically architected for \textit{generalizable multi-property optimization}. These models were trained on the \textit{MuMOInstruct} dataset, which is built to enhance generalization to complex in-domain and out-of-distribution optimization challenges~\cite{dey-etal-2025-mathtt}. In this paper, we employed the GeLLMO-P6-Mistral version.
\end{itemize}
\subsection*{Visualization analysis on BPQ}
\noindent For the BPQ task, we sampled four input molecules for visualization:
\begin{itemize}
    \item NCCCC[C@@H]1NC(=O)[C@H](CCCCN)NC1=O
    \item Nc1nc2c(=O)[nH]cnc2n1[C@@H]1O[C@H](CO)[C@@H](O)[C@H]1O
    \item CCOC(=O)[C@H](CCc1ccccc1)N[C@@H]1CCCN2CCC[C@H](C(=O)O)N2C1=O
    \item NC[C@@H]1O[C@H](C(=O)N2CCOCC2)[C@H](O)[C@H](O)[C@H]1O
\end{itemize}
For each input molecule, we generated 100 candidate molecules using our model for the BPQ optimization task. To visualize the distribution and diversity of the generated molecules, we computed Morgan fingerprints (radius = 2, 2048 bits) for both the original and generated molecules and projected them into a two-dimensional space using t-SNE. In the visualization, each point corresponds to a molecule, and the color intensity reflects the average score of BBBP, penalized LogP (pLogP), and QED. Lighter colors indicate higher overall property scores.
\subsection*{Details of docking analysis}
\noindent For the ClpP2 target, we used the binding pocket extracted by TamGen~\cite{wu2024tamgen}. We selected Syn-A003-01, a molecule generated by TamGen with an IC$_{50}$ of 20~$\mu$M, as the starting point for further optimization. We generated 4 candidate molecules with each model. Docking scores for all generated molecules were then computed using AutoDock Vina~\cite{trott2010autodock}.
\subsection*{Molecule optimization on real-world drug discovery cases}
\noindent We selected two real-world lead optimization tasks reported after October 2025~\cite{nie2025accelerated,meng2025discovery}. The first task focused on optimizing an allosteric inhibitor of MALT1~\cite{nie2025accelerated}. Following the original study, we used Compound 2, a lead molecule selected by FEP+, as the starting point. We aimed to improve its binding affinity for MALT1 by optimizing interactions within the small hydrophobic pocket around the gem-dimethyl substitution, while keeping the rest of the scaffold fixed. This process successfully yielded Compound 11. Subsequently, we performed an unconstrained optimization targeting kinetic solubility and permeability, ultimately obtaining the clinical candidate SGR-1505.

The second task focused on the optimization of CBL-B inhibitors~\cite{meng2025discovery}. We used Compound 8 as the starting point and followed a two-stage optimization strategy aligned with the structural insights presented in the original study. In the first stage, we applied our model to optimize Region B, corresponding to the triazole moiety, while keeping the rest of the molecular structure fixed. We obtained Compound 10 and performed further optimization by exploring substituents at the meta-position of the phenyl ring, a site identified as occupying a sterically permissive and chemically neutral region within the protein binding pocket.

For both stages, structural constraints were encoded in the property analysis. Molecule generation was conducted with a sampling temperature of 1, and 10,000 candidates were produced for each stage. The generated unique molecules were then selected and evaluated based on LogP, QED, Synthetic Accessibility (SA) scores, and molecular weight.

\subsection*{Case study on reaction ligand selection}
\noindent We used the dataset from Perera et al. \cite{perera2018platform}. To isolate the impact of ligand on catalytic efficiency, we organized the dataset into controlled reaction groups. Within each group, the reaction context—comprising the electrophile, nucleophile, palladium source, base, and solvent—was held constant, while the ligand identity and reaction yield varied. To facilitate physicochemical reasoning, each ligand was represented by its SMILES string augmented with computed descriptors: molecular weight (MW), partition coefficient (LogP), topological polar surface area (TPSA), and the number of rotatable bonds (RotBonds). These structured groupings served as the input for \ourF{} to decipher the mechanistic rationale connecting ligand features to reaction outcomes.


\section*{Acknowledgments}
\noindent We'd like to express our gratitude to Dr. Bin Feng and Dr. Hao Li from Peking University for their valuable guidance on the ChemCoTBench dataset. 
\bibliographystyle{naturemag}
\bibliography{sample}

\begin{thebibliography}{10}
\expandafter\ifx\csname url\endcsname\relax
  \def\url#1{\texttt{#1}}\fi
\expandafter\ifx\csname urlprefix\endcsname\relax\def\urlprefix{URL }\fi
\providecommand{\bibinfo}[2]{#2}
\providecommand{\eprint}[2][]{\url{#2}}

\bibitem{jiang2025deepseek}
\bibinfo{author}{Jiang, Q.}, \bibinfo{author}{Gao, Z.} \& \bibinfo{author}{Karniadakis, G.~E.}
\newblock \bibinfo{title}{Deepseek vs. chatgpt vs. claude: A comparative study for scientific computing and scientific machine learning tasks}.
\newblock \emph{\bibinfo{journal}{Theoretical and Applied Mechanics Letters}} \textbf{\bibinfo{volume}{15}}, \bibinfo{pages}{100583} (\bibinfo{year}{2025}).

\bibitem{zheng2025large}
\bibinfo{author}{Zheng, Y.} \emph{et~al.}
\newblock \bibinfo{title}{Large language models for scientific discovery in molecular property prediction}.
\newblock \emph{\bibinfo{journal}{Nature Machine Intelligence}} \bibinfo{pages}{1--11} (\bibinfo{year}{2025}).

\bibitem{dossetter2013matched}
\bibinfo{author}{Dossetter, A.~G.}, \bibinfo{author}{Griffen, E.~J.} \& \bibinfo{author}{Leach, A.~G.}
\newblock \bibinfo{title}{Matched molecular pair analysis in drug discovery}.
\newblock \emph{\bibinfo{journal}{Drug Discovery Today}} \textbf{\bibinfo{volume}{18}}, \bibinfo{pages}{724--731} (\bibinfo{year}{2013}).

\bibitem{Gioiello2020}
\bibinfo{author}{Gioiello, A.}, \bibinfo{author}{Piccinno, A.}, \bibinfo{author}{Lozza, A.~M.} \& \bibinfo{author}{Cerra, B.}
\newblock \bibinfo{title}{The medicinal chemistry in the era of machines and automation: Recent advances in continuous flow technology}.
\newblock \emph{\bibinfo{journal}{Journal of Medicinal Chemistry}} \textbf{\bibinfo{volume}{63}}, \bibinfo{pages}{6624--6647} (\bibinfo{year}{2020}).
\newblock \urlprefix\url{https://doi.org/10.1021/acs.jmedchem.9b01956}.

\bibitem{guo2025deepseek}
\bibinfo{author}{Guo, D.} \emph{et~al.}
\newblock \bibinfo{title}{Deepseek-r1: Incentivizing reasoning capability in llms via reinforcement learning}.
\newblock \emph{\bibinfo{journal}{arXiv preprint arXiv:2501.12948}}  (\bibinfo{year}{2025}).

\bibitem{jaech2024openai}
\bibinfo{author}{Jaech, A.} \emph{et~al.}
\newblock \bibinfo{title}{Openai o1 system card}.
\newblock \emph{\bibinfo{journal}{arXiv preprint arXiv:2412.16720}}  (\bibinfo{year}{2024}).

\bibitem{yang2025qwen3}
\bibinfo{author}{Yang, A.} \emph{et~al.}
\newblock \bibinfo{title}{Qwen3 technical report}.
\newblock \emph{\bibinfo{journal}{arXiv preprint arXiv:2505.09388}}  (\bibinfo{year}{2025}).

\bibitem{comanici2025gemini}
\bibinfo{author}{Comanici, G.} \emph{et~al.}
\newblock \bibinfo{title}{Gemini 2.5: Pushing the frontier with advanced reasoning, multimodality, long context, and next generation agentic capabilities}.
\newblock \emph{\bibinfo{journal}{arXiv preprint arXiv:2507.06261}}  (\bibinfo{year}{2025}).

\bibitem{Zheng2025}
\bibinfo{author}{Zheng, Y.} \emph{et~al.}
\newblock \bibinfo{title}{Large language models for drug discovery and development}.
\newblock \emph{\bibinfo{journal}{Patterns}} \textbf{\bibinfo{volume}{6}} (\bibinfo{year}{2025}).
\newblock \urlprefix\url{https://doi.org/10.1016/j.patter.2025.101346}.

\bibitem{wang2025medicalreasoningerallms}
\bibinfo{author}{Wang, W.} \emph{et~al.}
\newblock \bibinfo{title}{Medical reasoning in the era of llms: A systematic review of enhancement techniques and applications} (\bibinfo{year}{2025}).
\newblock \urlprefix\url{https://arxiv.org/abs/2508.00669}.
\newblock \eprint{2508.00669}.

\bibitem{liu2025chemauharnessreasoningllms}
\bibinfo{author}{Liu, X.} \emph{et~al.}
\newblock \bibinfo{title}{Chemau: Harness the reasoning of llms in chemical research with adaptive uncertainty estimation} (\bibinfo{year}{2025}).
\newblock \urlprefix\url{https://arxiv.org/abs/2506.01116}.
\newblock \eprint{2506.01116}.

\bibitem{erikawa2021mermaid}
\bibinfo{author}{Erikawa, D.}, \bibinfo{author}{Yasuo, N.} \& \bibinfo{author}{Sekijima, M.}
\newblock \bibinfo{title}{Mermaid: an open source automated hit-to-lead method based on deep reinforcement learning}.
\newblock \emph{\bibinfo{journal}{Journal of Cheminformatics}} \textbf{\bibinfo{volume}{13}}, \bibinfo{pages}{94} (\bibinfo{year}{2021}).

\bibitem{ahmed2025ai}
\bibinfo{author}{Ahmed, S.~A.} \emph{et~al.}
\newblock \bibinfo{title}{Ai-guided fragment-based drug design for virtual library screening and hit optimization}.
\newblock \emph{\bibinfo{journal}{Journal of Pharma Insights and Research}} \textbf{\bibinfo{volume}{3}}, \bibinfo{pages}{170--180} (\bibinfo{year}{2025}).

\bibitem{feng2025hierarchical}
\bibinfo{author}{Feng, B.} \emph{et~al.}
\newblock \bibinfo{title}{Hierarchical affinity landscape navigation through learning a shared pocket-ligand space}.
\newblock \emph{\bibinfo{journal}{Patterns}} \textbf{\bibinfo{volume}{6}} (\bibinfo{year}{2025}).

\bibitem{chakraborty2023utilizing}
\bibinfo{author}{Chakraborty, R.} \& \bibinfo{author}{Hasija, Y.}
\newblock \bibinfo{title}{Utilizing deep learning to explore chemical space for drug lead optimization}.
\newblock \emph{\bibinfo{journal}{Expert Systems with Applications}} \textbf{\bibinfo{volume}{229}}, \bibinfo{pages}{120592} (\bibinfo{year}{2023}).

\bibitem{sliwoski2014computational}
\bibinfo{author}{Sliwoski, G.}, \bibinfo{author}{Kothiwale, S.}, \bibinfo{author}{Meiler, J.} \& \bibinfo{author}{Lowe~Jr, E.~W.}
\newblock \bibinfo{title}{Computational methods in drug discovery}.
\newblock \emph{\bibinfo{journal}{Pharmacological reviews}} \textbf{\bibinfo{volume}{66}}, \bibinfo{pages}{334--395} (\bibinfo{year}{2014}).

\bibitem{elton2019deep}
\bibinfo{author}{Elton, D.~C.}, \bibinfo{author}{Boukouvalas, Z.}, \bibinfo{author}{Fuge, M.~D.} \& \bibinfo{author}{Chung, P.~W.}
\newblock \bibinfo{title}{Deep learning for molecular design—a review of the state of the art}.
\newblock \emph{\bibinfo{journal}{Molecular Systems Design \& Engineering}} \textbf{\bibinfo{volume}{4}}, \bibinfo{pages}{828--849} (\bibinfo{year}{2019}).

\bibitem{du2024machine}
\bibinfo{author}{Du, Y.} \emph{et~al.}
\newblock \bibinfo{title}{Machine learning-aided generative molecular design}.
\newblock \emph{\bibinfo{journal}{Nature Machine Intelligence}} \textbf{\bibinfo{volume}{6}}, \bibinfo{pages}{589--604} (\bibinfo{year}{2024}).

\bibitem{fromer2023computer}
\bibinfo{author}{Fromer, J.~C.} \& \bibinfo{author}{Coley, C.~W.}
\newblock \bibinfo{title}{Computer-aided multi-objective optimization in small molecule discovery}.
\newblock \emph{\bibinfo{journal}{Patterns}} \textbf{\bibinfo{volume}{4}} (\bibinfo{year}{2023}).

\bibitem{kim2023pubchem}
\bibinfo{author}{Kim, S.} \emph{et~al.}
\newblock \bibinfo{title}{Pubchem 2023 update}.
\newblock \emph{\bibinfo{journal}{Nucleic acids research}} \textbf{\bibinfo{volume}{51}}, \bibinfo{pages}{D1373--D1380} (\bibinfo{year}{2023}).

\bibitem{gaulton2012chembl}
\bibinfo{author}{Gaulton, A.} \emph{et~al.}
\newblock \bibinfo{title}{Chembl: a large-scale bioactivity database for drug discovery}.
\newblock \emph{\bibinfo{journal}{Nucleic acids research}} \textbf{\bibinfo{volume}{40}}, \bibinfo{pages}{D1100--D1107} (\bibinfo{year}{2012}).

\bibitem{wishart2006drugbank}
\bibinfo{author}{Wishart, D.~S.} \emph{et~al.}
\newblock \bibinfo{title}{Drugbank: a comprehensive resource for in silico drug discovery and exploration}.
\newblock \emph{\bibinfo{journal}{Nucleic acids research}} \textbf{\bibinfo{volume}{34}}, \bibinfo{pages}{D668--D672} (\bibinfo{year}{2006}).

\bibitem{jin2018junction}
\bibinfo{author}{Jin, W.}, \bibinfo{author}{Barzilay, R.} \& \bibinfo{author}{Jaakkola, T.}
\newblock \bibinfo{title}{Junction tree variational autoencoder for molecular graph generation}.
\newblock In \emph{\bibinfo{booktitle}{International conference on machine learning}}, \bibinfo{pages}{2323--2332} (\bibinfo{organization}{PMLR}, \bibinfo{year}{2018}).

\bibitem{fu2022reinforced}
\bibinfo{author}{Fu, T.}, \bibinfo{author}{Gao, W.}, \bibinfo{author}{Coley, C.} \& \bibinfo{author}{Sun, J.}
\newblock \bibinfo{title}{Reinforced genetic algorithm for structure-based drug design}.
\newblock \emph{\bibinfo{journal}{Advances in Neural Information Processing Systems}} \textbf{\bibinfo{volume}{35}}, \bibinfo{pages}{12325--12338} (\bibinfo{year}{2022}).

\bibitem{dey-etal-2025-mathtt}
\bibinfo{author}{Dey, V.}, \bibinfo{author}{Hu, X.} \& \bibinfo{author}{Ning, X.}
\newblock \bibinfo{title}{{GeLLM{\textthreesuperior}O}: Generalizing large language models for multi-property molecule optimization}.
\newblock In \bibinfo{editor}{Che, W.}, \bibinfo{editor}{Nabende, J.}, \bibinfo{editor}{Shutova, E.} \& \bibinfo{editor}{Pilehvar, M.~T.} (eds.) \emph{\bibinfo{booktitle}{Proceedings of the 63rd Annual Meeting of the Association for Computational Linguistics (Volume 1: Long Papers)}}, \bibinfo{pages}{25192--25221} (\bibinfo{publisher}{Association for Computational Linguistics}, \bibinfo{address}{Vienna, Austria}, \bibinfo{year}{2025}).
\newblock \urlprefix\url{https://aclanthology.org/2025.acl-long.1225/}.

\bibitem{zhu2025conditional}
\bibinfo{author}{Zhu, L.} \& \bibinfo{author}{Noutahi, E.}
\newblock \bibinfo{title}{Conditional chemical language models are versatile tools in drug discovery}.
\newblock \emph{\bibinfo{journal}{arXiv preprint arXiv:2507.10273}}  (\bibinfo{year}{2025}).

\bibitem{dey2025large}
\bibinfo{author}{Dey, V.}, \bibinfo{author}{Hu, X.} \& \bibinfo{author}{Ning, X.}
\newblock \bibinfo{title}{Large language models for controllable multi-property multi-objective molecule optimization}.
\newblock \emph{\bibinfo{journal}{arXiv preprint arXiv:2505.23987}}  (\bibinfo{year}{2025}).

\bibitem{wu2024leveraging}
\bibinfo{author}{Wu, Z.} \emph{et~al.}
\newblock \bibinfo{title}{Leveraging language model for advanced multiproperty molecular optimization via prompt engineering}.
\newblock \emph{\bibinfo{journal}{Nature Machine Intelligence}} \textbf{\bibinfo{volume}{6}}, \bibinfo{pages}{1359--1369} (\bibinfo{year}{2024}).

\bibitem{ye2025drugassist}
\bibinfo{author}{Ye, G.} \emph{et~al.}
\newblock \bibinfo{title}{Drugassist: A large language model for molecule optimization}.
\newblock \emph{\bibinfo{journal}{Briefings in Bioinformatics}} \textbf{\bibinfo{volume}{26}}, \bibinfo{pages}{bbae693} (\bibinfo{year}{2025}).

\bibitem{edwards2025mclm}
\bibinfo{author}{Edwards, C.} \emph{et~al.}
\newblock \bibinfo{title}{mclm: A function-infused and synthesis-friendly modular chemical language model}.
\newblock \emph{\bibinfo{journal}{arXiv preprint arXiv:2505.12565}}  (\bibinfo{year}{2025}).

\bibitem{liu2023multi}
\bibinfo{author}{Liu, S.} \emph{et~al.}
\newblock \bibinfo{title}{Multi-modal molecule structure--text model for text-based retrieval and editing}.
\newblock \emph{\bibinfo{journal}{Nature Machine Intelligence}} \textbf{\bibinfo{volume}{5}}, \bibinfo{pages}{1447--1457} (\bibinfo{year}{2023}).

\bibitem{dubey2024llama}
\bibinfo{author}{Dubey, A.} \emph{et~al.}
\newblock \bibinfo{title}{The llama 3 herd of models}.
\newblock \emph{\bibinfo{journal}{arXiv e-prints}} \bibinfo{pages}{arXiv--2407} (\bibinfo{year}{2024}).

\bibitem{wei2022cot}
\bibinfo{author}{Wei, J.} \emph{et~al.}
\newblock \bibinfo{title}{Chain-of-thought prompting elicits reasoning in large language models}.
\newblock In \bibinfo{editor}{Koyejo, S.} \emph{et~al.} (eds.) \emph{\bibinfo{booktitle}{Advances in Neural Information Processing Systems}}, vol.~\bibinfo{volume}{35}, \bibinfo{pages}{24824--24837} (\bibinfo{publisher}{Curran Associates, Inc.}, \bibinfo{year}{2022}).
\newblock \urlprefix\url{https://proceedings.neurips.cc/paper_files/paper/2022/file/9d5609613524ecf4f15af0f7b31abca4-Paper-Conference.pdf}.

\bibitem{gengnovo}
\bibinfo{author}{Geng, Z.} \emph{et~al.}
\newblock \bibinfo{title}{De novo molecular generation via connection-aware motif mining}.
\newblock In \emph{\bibinfo{booktitle}{The Eleventh International Conference on Learning Representations}}.

\bibitem{narayanan2025training}
\bibinfo{author}{Narayanan, S.~M.} \emph{et~al.}
\newblock \bibinfo{title}{Training a scientific reasoning model for chemistry}.
\newblock \emph{\bibinfo{journal}{arXiv preprint arXiv:2506.17238}}  (\bibinfo{year}{2025}).

\bibitem{yullasmol}
\bibinfo{author}{Yu, B.}, \bibinfo{author}{Baker, F.~N.}, \bibinfo{author}{Chen, Z.}, \bibinfo{author}{Ning, X.} \& \bibinfo{author}{Sun, H.}
\newblock \bibinfo{title}{Llasmol: Advancing large language models for chemistry with a large-scale, comprehensive, high-quality instruction tuning dataset}.
\newblock In \emph{\bibinfo{booktitle}{First Conference on Language Modeling}}.

\bibitem{maaten2008visualizing}
\bibinfo{author}{Maaten, L. v.~d.} \& \bibinfo{author}{Hinton, G.}
\newblock \bibinfo{title}{Visualizing data using t-sne}.
\newblock \emph{\bibinfo{journal}{Journal of machine learning research}} \textbf{\bibinfo{volume}{9}}, \bibinfo{pages}{2579--2605} (\bibinfo{year}{2008}).

\bibitem{nie2025accelerated}
\bibinfo{author}{Nie, Z.} \emph{et~al.}
\newblock \bibinfo{title}{Accelerated in silico discovery of sgr-1505: A potent malt1 allosteric inhibitor for the treatment of mature b-cell malignancies}.
\newblock \emph{\bibinfo{journal}{Journal of Medicinal Chemistry}} \textbf{\bibinfo{volume}{68}}, \bibinfo{pages}{23977--23992} (\bibinfo{year}{2025}).

\bibitem{meng2025discovery}
\bibinfo{author}{Meng, F.} \emph{et~al.}
\newblock \bibinfo{title}{Discovery and biological evaluation of novel, potent, and orally available cblb inhibitors}.
\newblock \emph{\bibinfo{journal}{Journal of Medicinal Chemistry}}  (\bibinfo{year}{2025}).

\bibitem{perera2018platform}
\bibinfo{author}{Perera, D.} \emph{et~al.}
\newblock \bibinfo{title}{A platform for automated nanomole-scale reaction screening and micromole-scale synthesis in flow}.
\newblock \emph{\bibinfo{journal}{Science}} \textbf{\bibinfo{volume}{359}}, \bibinfo{pages}{429--434} (\bibinfo{year}{2018}).

\bibitem{wang2025scireasoner}
\bibinfo{author}{Wang, Y.} \emph{et~al.}
\newblock \bibinfo{title}{Scireasoner: Laying the scientific reasoning ground across disciplines}.
\newblock \emph{\bibinfo{journal}{arXiv preprint arXiv:2509.21320}}  (\bibinfo{year}{2025}).

\bibitem{fallahpour2025bioreason}
\bibinfo{author}{Fallahpour, A.} \emph{et~al.}
\newblock \bibinfo{title}{Bioreason: Incentivizing multimodal biological reasoning within a dna-llm model}.
\newblock \emph{\bibinfo{journal}{arXiv preprint arXiv:2505.23579}}  (\bibinfo{year}{2025}).

\bibitem{wang2025chem}
\bibinfo{author}{Wang, W.} \emph{et~al.}
\newblock \bibinfo{title}{Chem-r: Learning to reason as a chemist}.
\newblock \emph{\bibinfo{journal}{arXiv preprint arXiv:2510.16880}}  (\bibinfo{year}{2025}).

\bibitem{zhao2025chemdfm}
\bibinfo{author}{Zhao, Z.} \emph{et~al.}
\newblock \bibinfo{title}{Chemdfm-r: A chemical reasoning llm enhanced with atomized chemical knowledge}.
\newblock \emph{\bibinfo{journal}{arXiv preprint arXiv:2507.21990}}  (\bibinfo{year}{2025}).

\bibitem{boiko2023autonomous}
\bibinfo{author}{Boiko, D.~A.}, \bibinfo{author}{MacKnight, R.}, \bibinfo{author}{Kline, B.} \& \bibinfo{author}{Gomes, G.}
\newblock \bibinfo{title}{Autonomous chemical research with large language models}.
\newblock \emph{\bibinfo{journal}{Nature}} \textbf{\bibinfo{volume}{624}}, \bibinfo{pages}{570--578} (\bibinfo{year}{2023}).

\bibitem{penades2025ai}
\bibinfo{author}{Penad{\'e}s, J.~R.} \emph{et~al.}
\newblock \bibinfo{title}{Ai mirrors experimental science to uncover a mechanism of gene transfer crucial to bacterial evolution}.
\newblock \emph{\bibinfo{journal}{Cell}} \textbf{\bibinfo{volume}{188}}, \bibinfo{pages}{6654--6665} (\bibinfo{year}{2025}).

\bibitem{graff2021accelerating}
\bibinfo{author}{Graff, D.~E.}, \bibinfo{author}{Shakhnovich, E.~I.} \& \bibinfo{author}{Coley, C.~W.}
\newblock \bibinfo{title}{Accelerating high-throughput virtual screening through molecular pool-based active learning}.
\newblock \emph{\bibinfo{journal}{Chemical science}} \textbf{\bibinfo{volume}{12}}, \bibinfo{pages}{7866--7881} (\bibinfo{year}{2021}).

\bibitem{korovina2020chembo}
\bibinfo{author}{Korovina, K.} \emph{et~al.}
\newblock \bibinfo{title}{Chembo: Bayesian optimization of small organic molecules with synthesizable recommendations}.
\newblock In \emph{\bibinfo{booktitle}{International Conference on Artificial Intelligence and Statistics}}, \bibinfo{pages}{3393--3403} (\bibinfo{organization}{PMLR}, \bibinfo{year}{2020}).

\bibitem{yoshikawa2018population}
\bibinfo{author}{Yoshikawa, N.} \emph{et~al.}
\newblock \bibinfo{title}{Population-based de novo molecule generation, using grammatical evolution}.
\newblock \emph{\bibinfo{journal}{Chemistry Letters}} \textbf{\bibinfo{volume}{47}}, \bibinfo{pages}{1431--1434} (\bibinfo{year}{2018}).

\bibitem{nigamaugmenting}
\bibinfo{author}{Nigam, A.}, \bibinfo{author}{Friederich, P.}, \bibinfo{author}{Krenn, M.} \& \bibinfo{author}{Aspuru-Guzik, A.}
\newblock \bibinfo{title}{Augmenting genetic algorithms with deep neural networks for exploring the chemical space}.
\newblock In \emph{\bibinfo{booktitle}{International Conference on Learning Representations}}.

\bibitem{nigam2021beyond}
\bibinfo{author}{Nigam, A.}, \bibinfo{author}{Pollice, R.}, \bibinfo{author}{Krenn, M.}, \bibinfo{author}{dos Passos~Gomes, G.} \& \bibinfo{author}{Aspuru-Guzik, A.}
\newblock \bibinfo{title}{Beyond generative models: superfast traversal, optimization, novelty, exploration and discovery (stoned) algorithm for molecules using selfies}.
\newblock \emph{\bibinfo{journal}{Chemical science}} \textbf{\bibinfo{volume}{12}}, \bibinfo{pages}{7079--7090} (\bibinfo{year}{2021}).

\bibitem{jensen2019graph}
\bibinfo{author}{Jensen, J.~H.}
\newblock \bibinfo{title}{A graph-based genetic algorithm and generative model/monte carlo tree search for the exploration of chemical space}.
\newblock \emph{\bibinfo{journal}{Chemical science}} \textbf{\bibinfo{volume}{10}}, \bibinfo{pages}{3567--3572} (\bibinfo{year}{2019}).

\bibitem{gaoamortized}
\bibinfo{author}{Gao, W.}, \bibinfo{author}{Mercado, R.} \& \bibinfo{author}{Coley, C.~W.}
\newblock \bibinfo{title}{Amortized tree generation for bottom-up synthesis planning and synthesizable molecular design}.
\newblock In \emph{\bibinfo{booktitle}{International Conference on Learning Representations}}.

\bibitem{gao2024generative}
\bibinfo{author}{Gao, W.}, \bibinfo{author}{Luo, S.} \& \bibinfo{author}{Coley, C.~W.}
\newblock \bibinfo{title}{Generative artificial intelligence for navigating synthesizable chemical space}.
\newblock \emph{\bibinfo{journal}{arXiv preprint arXiv:2410.03494}}  (\bibinfo{year}{2024}).

\bibitem{maus2022local}
\bibinfo{author}{Maus, N.} \emph{et~al.}
\newblock \bibinfo{title}{Local latent space bayesian optimization over structured inputs}.
\newblock \emph{\bibinfo{journal}{Advances in neural information processing systems}} \textbf{\bibinfo{volume}{35}}, \bibinfo{pages}{34505--34518} (\bibinfo{year}{2022}).

\bibitem{de2018molgan}
\bibinfo{author}{De~Cao, N.} \& \bibinfo{author}{Kipf, T.}
\newblock \bibinfo{title}{Molgan: An implicit generative model for small molecular graphs}.
\newblock \emph{\bibinfo{journal}{arXiv preprint arXiv:1805.11973}}  (\bibinfo{year}{2018}).

\bibitem{xiemars}
\bibinfo{author}{Xie, Y.} \emph{et~al.}
\newblock \bibinfo{title}{Mars: Markov molecular sampling for multi-objective drug discovery}.
\newblock In \emph{\bibinfo{booktitle}{International Conference on Learning Representations}}.

\bibitem{lee2023exploring}
\bibinfo{author}{Lee, S.}, \bibinfo{author}{Jo, J.} \& \bibinfo{author}{Hwang, S.~J.}
\newblock \bibinfo{title}{Exploring chemical space with score-based out-of-distribution generation}.
\newblock In \emph{\bibinfo{booktitle}{International Conference on Machine Learning}}, \bibinfo{pages}{18872--18892} (\bibinfo{organization}{PMLR}, \bibinfo{year}{2023}).

\bibitem{lee2024molecule}
\bibinfo{author}{Lee, S.} \emph{et~al.}
\newblock \bibinfo{title}{Molecule generation with fragment retrieval augmentation}.
\newblock \emph{\bibinfo{journal}{Advances in Neural Information Processing Systems}} \textbf{\bibinfo{volume}{37}}, \bibinfo{pages}{132463--132490} (\bibinfo{year}{2024}).

\bibitem{chen2025uncertainty}
\bibinfo{author}{Chen, L.-Y.} \& \bibinfo{author}{Li, Y.-P.}
\newblock \bibinfo{title}{Uncertainty quantification with graph neural networks for efficient molecular design}.
\newblock \emph{\bibinfo{journal}{Nature Communications}} \textbf{\bibinfo{volume}{16}}, \bibinfo{pages}{3262} (\bibinfo{year}{2025}).

\bibitem{zhu2023sample}
\bibinfo{author}{Zhu, Y.} \emph{et~al.}
\newblock \bibinfo{title}{Sample-efficient multi-objective molecular optimization with gflownets}.
\newblock \emph{\bibinfo{journal}{Advances in Neural Information Processing Systems}} \textbf{\bibinfo{volume}{36}}, \bibinfo{pages}{79667--79684} (\bibinfo{year}{2023}).

\bibitem{liu2024graph}
\bibinfo{author}{Liu, G.}, \bibinfo{author}{Xu, J.}, \bibinfo{author}{Luo, T.} \& \bibinfo{author}{Jiang, M.}
\newblock \bibinfo{title}{Graph diffusion transformers for multi-conditional molecular generation}.
\newblock \emph{\bibinfo{journal}{Advances in Neural Information Processing Systems}} \textbf{\bibinfo{volume}{37}}, \bibinfo{pages}{8065--8092} (\bibinfo{year}{2024}).

\bibitem{lee2025genmol}
\bibinfo{author}{Lee, S.} \emph{et~al.}
\newblock \bibinfo{title}{Genmol: A drug discovery generalist with discrete diffusion}.
\newblock \emph{\bibinfo{journal}{arXiv preprint arXiv:2501.06158}}  (\bibinfo{year}{2025}).

\bibitem{he2021molecular}
\bibinfo{author}{He, J.} \emph{et~al.}
\newblock \bibinfo{title}{Molecular optimization by capturing chemist’s intuition using deep neural networks}.
\newblock \emph{\bibinfo{journal}{Journal of cheminformatics}} \textbf{\bibinfo{volume}{13}}, \bibinfo{pages}{26} (\bibinfo{year}{2021}).

\bibitem{fu2021mimosa}
\bibinfo{author}{Fu, T.}, \bibinfo{author}{Xiao, C.}, \bibinfo{author}{Li, X.}, \bibinfo{author}{Glass, L.~M.} \& \bibinfo{author}{Sun, J.}
\newblock \bibinfo{title}{Mimosa: Multi-constraint molecule sampling for molecule optimization}.
\newblock In \emph{\bibinfo{booktitle}{Proceedings of the AAAI Conference on Artificial Intelligence}}, vol.~\bibinfo{volume}{35}, \bibinfo{pages}{125--133} (\bibinfo{year}{2021}).

\bibitem{brown2019guacamol}
\bibinfo{author}{Brown, N.}, \bibinfo{author}{Fiscato, M.}, \bibinfo{author}{Segler, M.~H.} \& \bibinfo{author}{Vaucher, A.~C.}
\newblock \bibinfo{title}{Guacamol: benchmarking models for de novo molecular design}.
\newblock \emph{\bibinfo{journal}{Journal of chemical information and modeling}} \textbf{\bibinfo{volume}{59}}, \bibinfo{pages}{1096--1108} (\bibinfo{year}{2019}).

\bibitem{olivecrona2017molecular}
\bibinfo{author}{Olivecrona, M.}, \bibinfo{author}{Blaschke, T.}, \bibinfo{author}{Engkvist, O.} \& \bibinfo{author}{Chen, H.}
\newblock \bibinfo{title}{Molecular de-novo design through deep reinforcement learning}.
\newblock \emph{\bibinfo{journal}{Journal of cheminformatics}} \textbf{\bibinfo{volume}{9}}, \bibinfo{pages}{48} (\bibinfo{year}{2017}).

\bibitem{zhou2019optimization}
\bibinfo{author}{Zhou, Z.}, \bibinfo{author}{Kearnes, S.}, \bibinfo{author}{Li, L.}, \bibinfo{author}{Zare, R.~N.} \& \bibinfo{author}{Riley, P.}
\newblock \bibinfo{title}{Optimization of molecules via deep reinforcement learning}.
\newblock \emph{\bibinfo{journal}{Scientific reports}} \textbf{\bibinfo{volume}{9}}, \bibinfo{pages}{10752} (\bibinfo{year}{2019}).

\bibitem{you2018graph}
\bibinfo{author}{You, J.}, \bibinfo{author}{Liu, B.}, \bibinfo{author}{Ying, Z.}, \bibinfo{author}{Pande, V.} \& \bibinfo{author}{Leskovec, J.}
\newblock \bibinfo{title}{Graph convolutional policy network for goal-directed molecular graph generation}.
\newblock \emph{\bibinfo{journal}{Advances in neural information processing systems}} \textbf{\bibinfo{volume}{31}} (\bibinfo{year}{2018}).

\bibitem{jin2020multi}
\bibinfo{author}{Jin, W.}, \bibinfo{author}{Barzilay, R.} \& \bibinfo{author}{Jaakkola, T.}
\newblock \bibinfo{title}{Multi-objective molecule generation using interpretable substructures}.
\newblock In \emph{\bibinfo{booktitle}{International conference on machine learning}}, \bibinfo{pages}{4849--4859} (\bibinfo{organization}{PMLR}, \bibinfo{year}{2020}).

\bibitem{yang2021hit}
\bibinfo{author}{Yang, S.}, \bibinfo{author}{Hwang, D.}, \bibinfo{author}{Lee, S.}, \bibinfo{author}{Ryu, S.} \& \bibinfo{author}{Hwang, S.~J.}
\newblock \bibinfo{title}{Hit and lead discovery with explorative rl and fragment-based molecule generation}.
\newblock \emph{\bibinfo{journal}{Advances in Neural Information Processing Systems}} \textbf{\bibinfo{volume}{34}}, \bibinfo{pages}{7924--7936} (\bibinfo{year}{2021}).

\bibitem{horwood2020molecular}
\bibinfo{author}{Horwood, J.} \& \bibinfo{author}{Noutahi, E.}
\newblock \bibinfo{title}{Molecular design in synthetically accessible chemical space via deep reinforcement learning}.
\newblock \emph{\bibinfo{journal}{ACS omega}} \textbf{\bibinfo{volume}{5}}, \bibinfo{pages}{32984--32994} (\bibinfo{year}{2020}).

\bibitem{dodds2024sample}
\bibinfo{author}{Dodds, M.} \emph{et~al.}
\newblock \bibinfo{title}{Sample efficient reinforcement learning with active learning for molecular design}.
\newblock \emph{\bibinfo{journal}{Chemical Science}} \textbf{\bibinfo{volume}{15}}, \bibinfo{pages}{4146--4160} (\bibinfo{year}{2024}).

\bibitem{bou2024acegen}
\bibinfo{author}{Bou, A.} \emph{et~al.}
\newblock \bibinfo{title}{Acegen: Reinforcement learning of generative chemical agents for drug discovery}.
\newblock \emph{\bibinfo{journal}{Journal of Chemical Information and Modeling}} \textbf{\bibinfo{volume}{64}}, \bibinfo{pages}{5900--5911} (\bibinfo{year}{2024}).

\bibitem{guo2025directly}
\bibinfo{author}{Guo, J.} \& \bibinfo{author}{Schwaller, P.}
\newblock \bibinfo{title}{Directly optimizing for synthesizability in generative molecular design using retrosynthesis models}.
\newblock \emph{\bibinfo{journal}{Chemical Science}} \textbf{\bibinfo{volume}{16}}, \bibinfo{pages}{6943--6956} (\bibinfo{year}{2025}).

\bibitem{yoshizawa2025data}
\bibinfo{author}{Yoshizawa, T.} \emph{et~al.}
\newblock \bibinfo{title}{A data-driven generative strategy to avoid reward hacking in multi-objective molecular design}.
\newblock \emph{\bibinfo{journal}{Nature Communications}} \textbf{\bibinfo{volume}{16}}, \bibinfo{pages}{2409} (\bibinfo{year}{2025}).

\bibitem{shen2021deep}
\bibinfo{author}{Shen, C.}, \bibinfo{author}{Krenn, M.}, \bibinfo{author}{Eppel, S.} \& \bibinfo{author}{Aspuru-Guzik, A.}
\newblock \bibinfo{title}{Deep molecular dreaming: inverse machine learning for de-novo molecular design and interpretability with surjective representations}.
\newblock \emph{\bibinfo{journal}{Machine Learning: Science and Technology}} \textbf{\bibinfo{volume}{2}}, \bibinfo{pages}{03LT02} (\bibinfo{year}{2021}).

\bibitem{fudifferentiable}
\bibinfo{author}{Fu, T.} \emph{et~al.}
\newblock \bibinfo{title}{Differentiable scaffolding tree for molecule optimization}.
\newblock In \emph{\bibinfo{booktitle}{International Conference on Learning Representations}}.

\bibitem{niuinversiongnn}
\bibinfo{author}{Niu, Y.} \emph{et~al.}
\newblock \bibinfo{title}{Inversiongnn: A dual path network for multi-property molecular optimization}.
\newblock In \emph{\bibinfo{booktitle}{The Thirteenth International Conference on Learning Representations}}.

\bibitem{loeffler2024reinvent}
\bibinfo{author}{Loeffler, H.~H.} \emph{et~al.}
\newblock \bibinfo{title}{Reinvent 4: modern ai--driven generative molecule design}.
\newblock \emph{\bibinfo{journal}{Journal of Cheminformatics}} \textbf{\bibinfo{volume}{16}}, \bibinfo{pages}{20} (\bibinfo{year}{2024}).

\bibitem{wang2025polo}
\bibinfo{author}{Wang, Z.} \emph{et~al.}
\newblock \bibinfo{title}{Polo: Preference-guided multi-turn reinforcement learning for lead optimization}.
\newblock \emph{\bibinfo{journal}{arXiv preprint arXiv:2509.21737}}  (\bibinfo{year}{2025}).

\bibitem{xia2025nature}
\bibinfo{author}{Xia, Y.} \emph{et~al.}
\newblock \bibinfo{title}{Nature language model: Deciphering the language of nature for scientific discovery}.
\newblock \emph{\bibinfo{journal}{arXiv preprint arXiv:2502.07527}}  (\bibinfo{year}{2025}).

\bibitem{wangefficient}
\bibinfo{author}{Wang, H.} \emph{et~al.}
\newblock \bibinfo{title}{Efficient evolutionary search over chemical space with large language models}.
\newblock In \emph{\bibinfo{booktitle}{The Thirteenth International Conference on Learning Representations}}.

\bibitem{nguyenlico}
\bibinfo{author}{Nguyen, T.} \& \bibinfo{author}{Grover, A.}
\newblock \bibinfo{title}{Lico: Large language models for in-context molecular optimization}.
\newblock In \emph{\bibinfo{booktitle}{The Thirteenth International Conference on Learning Representations}}.

\bibitem{xia2024evolutionary}
\bibinfo{author}{Xia, X.} \emph{et~al.}
\newblock \bibinfo{title}{Evolutionary multiobjective molecule optimization in an implicit chemical space}.
\newblock \emph{\bibinfo{journal}{Journal of Chemical Information and Modeling}} \textbf{\bibinfo{volume}{64}}, \bibinfo{pages}{5161--5174} (\bibinfo{year}{2024}).

\bibitem{liu2023chatgpt}
\bibinfo{author}{Liu, S.} \emph{et~al.}
\newblock \bibinfo{title}{Chatgpt-powered conversational drug editing using retrieval and domain feedback}.
\newblock \emph{\bibinfo{journal}{arXiv preprint arXiv:2305.18090}}  (\bibinfo{year}{2023}).

\bibitem{kim2025mt}
\bibinfo{author}{Kim, H.}, \bibinfo{author}{Jang, Y.} \& \bibinfo{author}{Ahn, S.}
\newblock \bibinfo{title}{Mt-mol: Multi agent system with tool-based reasoning for molecular optimization}.
\newblock \emph{\bibinfo{journal}{arXiv preprint arXiv:2505.20820}}  (\bibinfo{year}{2025}).

\bibitem{gao2025pharmagents}
\bibinfo{author}{Gao, B.} \emph{et~al.}
\newblock \bibinfo{title}{Pharmagents: Building a virtual pharma with large language model agents}.
\newblock \emph{\bibinfo{journal}{arXiv preprint arXiv:2503.22164}}  (\bibinfo{year}{2025}).

\bibitem{yu2025collaborative}
\bibinfo{author}{Yu, J.} \emph{et~al.}
\newblock \bibinfo{title}{Collaborative expert llms guided multi-objective molecular optimization}.
\newblock \emph{\bibinfo{journal}{arXiv preprint arXiv:2503.03503}}  (\bibinfo{year}{2025}).

\bibitem{zhu2025coder}
\bibinfo{author}{Zhu, W.} \emph{et~al.}
\newblock \bibinfo{title}{Coder as editor: Code-driven interpretable molecular optimization}.
\newblock \emph{\bibinfo{journal}{arXiv preprint arXiv:2510.14455}}  (\bibinfo{year}{2025}).

\bibitem{loshchilovdecoupled}
\bibinfo{author}{Loshchilov, I.} \& \bibinfo{author}{Hutter, F.}
\newblock \bibinfo{title}{Decoupled weight decay regularization}.
\newblock In \emph{\bibinfo{booktitle}{International Conference on Learning Representations}}.

\bibitem{li2025chemicalqaevaluatingllms}
\bibinfo{author}{Li, H.} \emph{et~al.}
\newblock \bibinfo{title}{Beyond chemical qa: Evaluating llm's chemical reasoning with modular chemical operations} (\bibinfo{year}{2025}).
\newblock \urlprefix\url{https://arxiv.org/abs/2505.21318}.
\newblock \eprint{2505.21318}.

\bibitem{Huang2021tdc}
\bibinfo{author}{Huang, K.} \emph{et~al.}
\newblock \bibinfo{title}{Therapeutics data commons: Machine learning datasets and tasks for drug discovery and development}.
\newblock \emph{\bibinfo{journal}{Proceedings of Neural Information Processing Systems, NeurIPS Datasets and Benchmarks}}  (\bibinfo{year}{2021}).

\bibitem{10.1093/nar/gks1111}
\bibinfo{author}{Yang, W.} \emph{et~al.}
\newblock \bibinfo{title}{Genomics of drug sensitivity in cancer (gdsc): a resource for therapeutic biomarker discovery in cancer cells}.
\newblock \emph{\bibinfo{journal}{Nucleic Acids Research}} \textbf{\bibinfo{volume}{41}}, \bibinfo{pages}{D955--D961} (\bibinfo{year}{2012}).
\newblock \urlprefix\url{https://doi.org/10.1093/nar/gks1111}.
\newblock \eprint{https://academic.oup.com/nar/article-pdf/41/D1/D955/3626591/gks1111.pdf}.

\bibitem{Swanson2024-qb}
\bibinfo{author}{Swanson, K.} \emph{et~al.}
\newblock \bibinfo{title}{{ADMET-AI}: a machine learning {ADMET} platform for evaluation of large-scale chemical libraries}.
\newblock \emph{\bibinfo{journal}{Bioinformatics}} \textbf{\bibinfo{volume}{40}} (\bibinfo{year}{2024}).

\bibitem{feng2024bioactivity}
\bibinfo{author}{Feng, B.} \emph{et~al.}
\newblock \bibinfo{title}{A bioactivity foundation model using pairwise meta-learning}.
\newblock \emph{\bibinfo{journal}{Nature Machine Intelligence}} \textbf{\bibinfo{volume}{6}}, \bibinfo{pages}{962--974} (\bibinfo{year}{2024}).

\bibitem{wu2024tamgen}
\bibinfo{author}{Wu, K.} \emph{et~al.}
\newblock \bibinfo{title}{Tamgen: drug design with target-aware molecule generation through a chemical language model}.
\newblock \emph{\bibinfo{journal}{Nature Communications}} \textbf{\bibinfo{volume}{15}}, \bibinfo{pages}{9360} (\bibinfo{year}{2024}).

\bibitem{trott2010autodock}
\bibinfo{author}{Trott, O.} \& \bibinfo{author}{Olson, A.~J.}
\newblock \bibinfo{title}{Autodock vina: improving the speed and accuracy of docking with a new scoring function, efficient optimization, and multithreading}.
\newblock \emph{\bibinfo{journal}{Journal of computational chemistry}} \textbf{\bibinfo{volume}{31}}, \bibinfo{pages}{455--461} (\bibinfo{year}{2010}).

\end{thebibliography}

\newpage
\appendix
\setcounter{figure}{0}  
\renewcommand{\thefigure}{S\arabic{figure}}
\setcounter{table}{0}  
\renewcommand{\thetable}{S\arabic{table}}
\section*{Supplementary Information}
\subsection*{Supplementary figures}
\begin{mdframed}[backgroundcolor=white, linecolor=black, linewidth=1pt]
Assume you are a very experienced chemist. The following data includes several groups of molecules and their corresponding property value. 

The property is: \{property\}

\{higher/lower\} value means better.

In each group, the molecules share a common structure, and each molecule has its unique structure.

[*:X] means the connect point between common structure and unique structure.

\{groups\}

Please infer step-by-step: 

1. analysis the chemical principles in the target property (\{property\})

2. compare unique structures within groups, consider the molecules with significant different values

3. come up with detailed chemical rules of a molecule to predict the property. 

4. Finally, output a JSON to list the analysis in 1. and rules in 3., the output should be in the format of 

\{``property analysis": analysis, ``rules": [\{``chemical rule": rule1, ``molecule evidence": [[[group x1, molecule y1], [group x1, molecule y2]], [[group x2, molecule y1], [group x2, molecule y2]],...]\}, ...]\}

A rule should be multiple sentences contain the property/structure related to the prediction and the chemical mechanism behind it.
A molecule evidence should be a pair of molecules in the same group with different property value, which reflect the rule.
\end{mdframed}  
\captionof{figure}{Prompt for rationale induction} \label{fig:prompt}
\newpage
\begin{figure}[!htp]
    \centering
    \includegraphics[width=\linewidth]{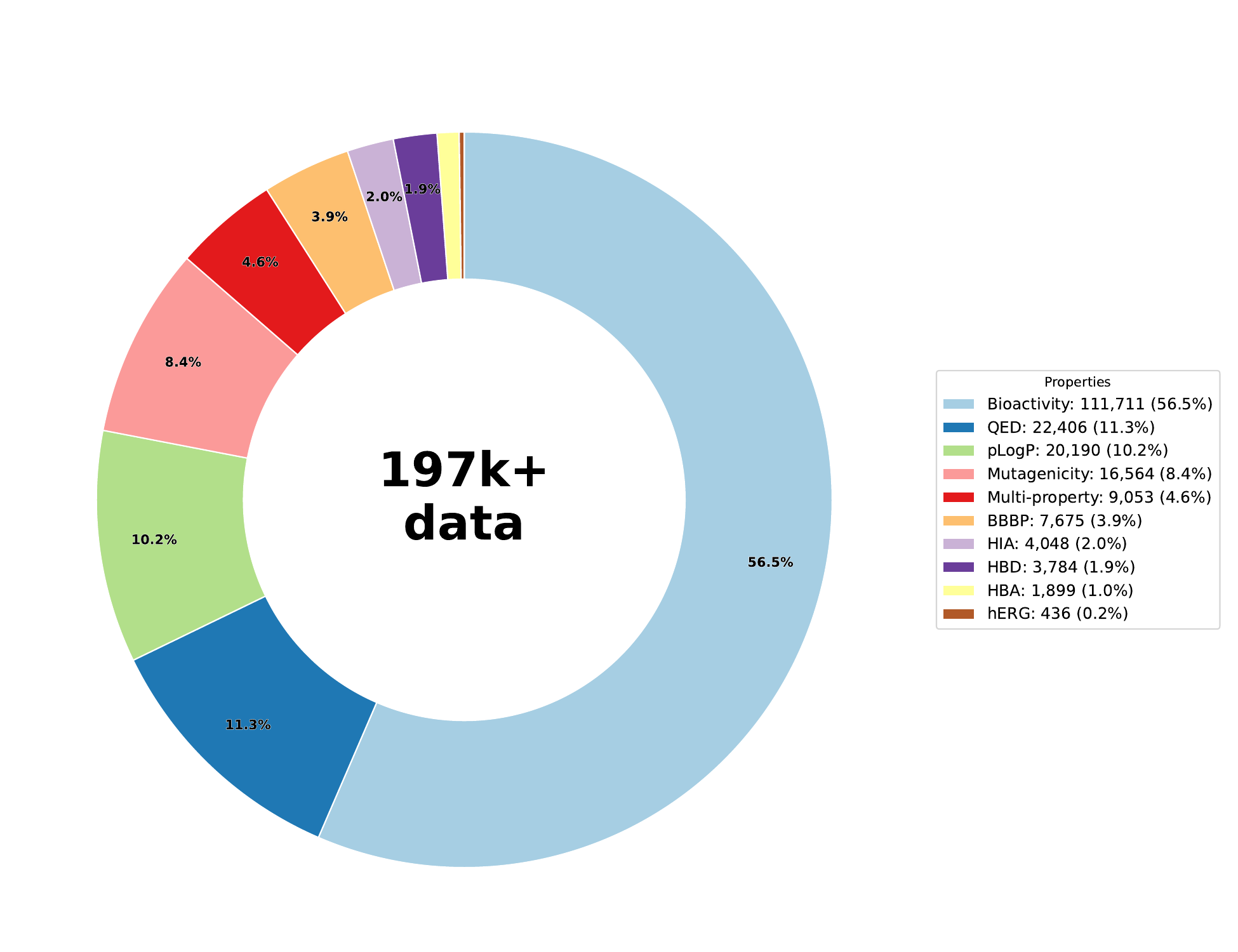} 
    \caption{Data distribution of the 197k+ molecule optimization reasoning processes. The pie chart categorizes the collected optimization paths by their target properties, spanning various single-property tasks (e.g., Bioactivity, QED, ADMET) and multi-property combinations.}
    \label{fig:data_stat}
\end{figure}
\newpage
\begin{mdframed}[backgroundcolor=white, linecolor=black, linewidth=1pt]
\textbf{Instruction:} For Fatty-acid amide hydrolase 1, propose modifications to <mol>O=C(Nc\\1ccccc1)N1CCN(Cc2ccc(Cl)cc2)CC1</mol> to develop an optimized molecule.  Let's think step by step. \\
\textbf{Answer: }<property\_analysis>  The property is binding affinity with Fatty-acid amide hydrolase 1. The inhibition of human FAAH (Fatty Acid Amide Hydrolase) is influenced by the molecular interactions between the inhibitor and the active site of the enzyme. Key chemical principles affecting this property include the presence of specific functional groups that can engage in hydrogen bonding, hydrophobic interactions, and electronic interactions. Electron-withdrawing groups such as fluorine, chlorine, and nitrile groups can enhance binding affinity by modifying the electronic distribution, thus improving interactions with the enzyme. Steric factors also play a role, where bulkier groups may either enhance specificity through better fitting into the active site pockets or hinder binding if they cause steric clashes. Additionally, the presence of aromatic rings can facilitate pi-pi stacking interactions with aromatic residues in the enzyme’s active site, contributing to stronger inhibition. </property\_analysis> <rule> Presence of multiple fluorine atoms significantly enhances FAAH inhibition due to increased electron-withdrawing effect and potential for stronger hydrophobic interactions. </rule> <rule> Electron-withdrawing groups at the connection point enhance FAAH inhibition by increasing binding affinity through electronic interactions. </rule> <modification> So we should keep <fragA>O=C(Nc1ccccc1)N1CCN(Cc2ccc([:100000])cc2)CC1</fragA>, and modify <fragB>Cl[:100000]</fragB>. </modification> The optimized fragment is <fragB>FC(F)(F)[*:100000]</fragB>
\end{mdframed}  
\captionof{figure}{An example for the reasoning data} \label{fig:example}
\newpage
\begin{figure}[!htp]
    \centering
    \includegraphics[width=0.5\linewidth]{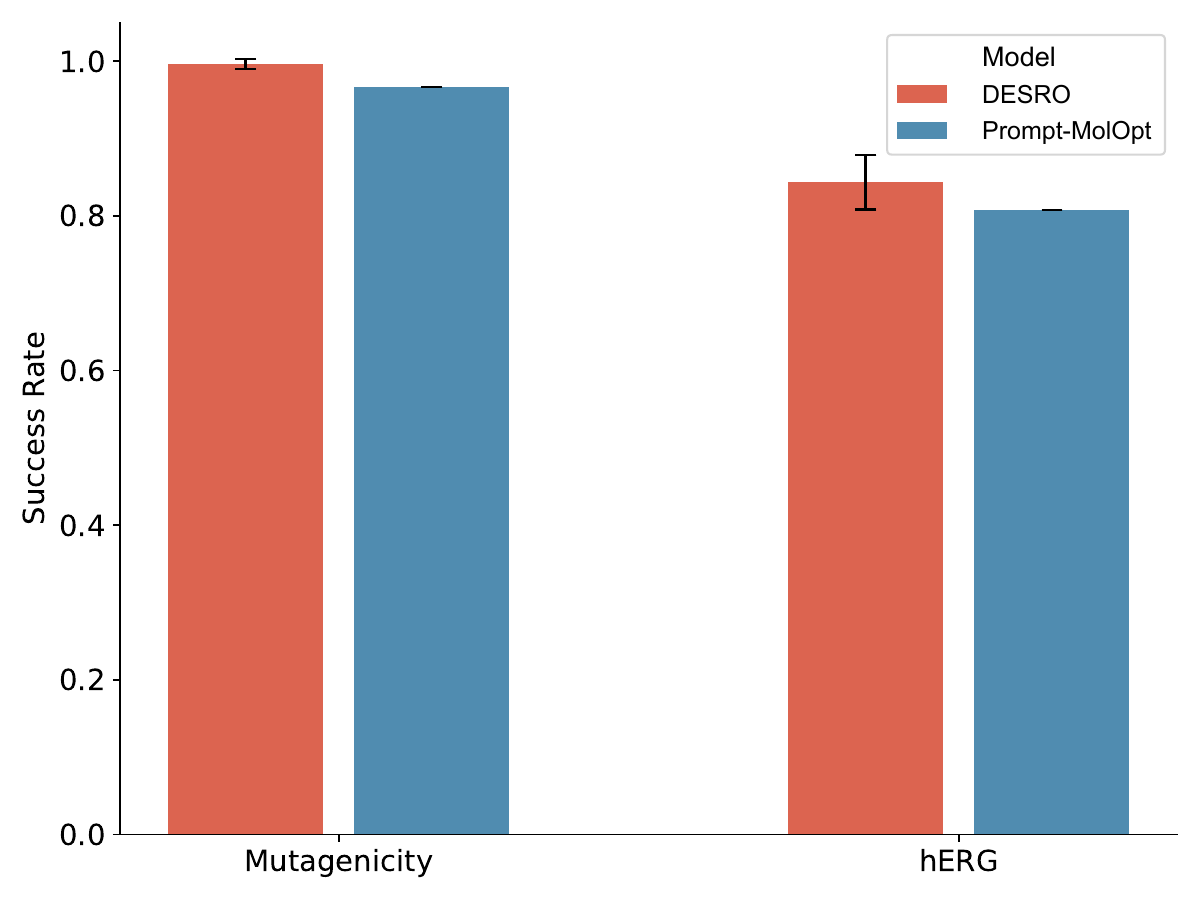} 
    \caption{Bar plot comparing \ourM{} and Prompt-MolOpt on Mutagenicity and hERG in terms of Success Rate.}
    \label{fig:prompt-molopt}
\end{figure}
\begin{figure}[!htp]
    \centering
    \includegraphics[width=0.5\linewidth]{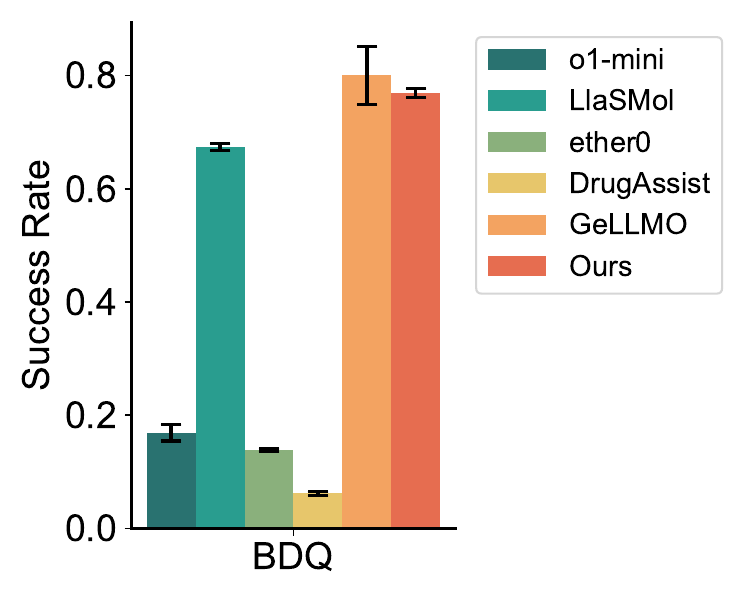} 
    \caption{Bar plot comparing \ourM{} and baselines on BDQ (increase BBBP, DRD2 and QED) in terms of Success Rate.}
    \label{fig:bdq}
\end{figure}
\newpage
\begin{figure}[!htp]
    \centering
    \includegraphics[width=\linewidth]{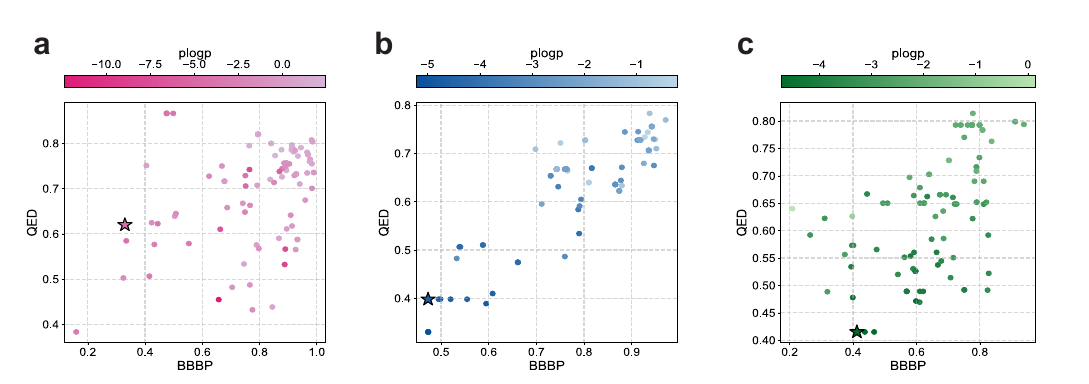} 
    \caption{\textbf{a}-\textbf{c}. Scatter plots illustrating the property distribution of generated molecules for three selected starting scaffolds in the BPQ task in terms of QED and BBBP, colored by pLogP values. }
    \label{fig:scatter}
\end{figure}
\begin{figure}[!htp]
    \centering
    \includegraphics[width=0.5\linewidth]{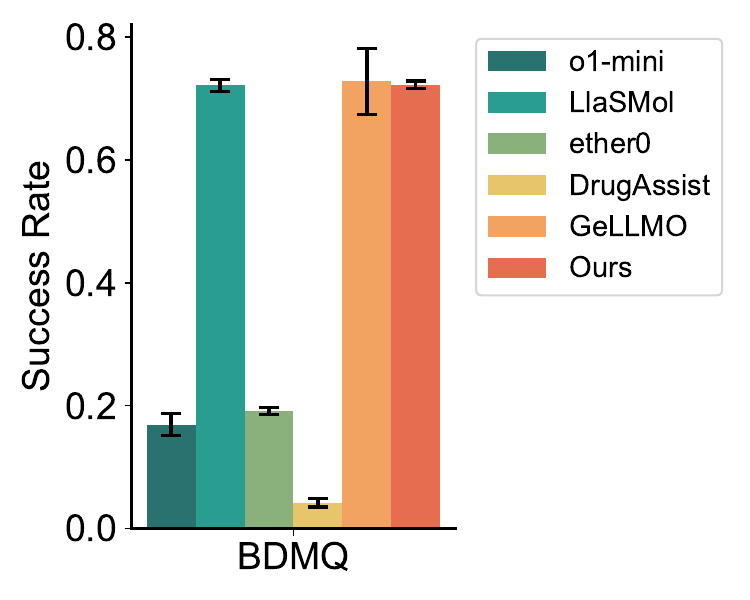} 
    \caption{Bar plot comparing the success rates of {\ourM} and baselines on the OOD multi-property optimization tasks, BDMQ (increase BBBP, DRD2, QED, and decrease mutagenicity).}
    \label{fig:ood_multi}
\end{figure}
\newpage
\begin{figure}[!htp]
    \centering
    \includegraphics[width=0.8\linewidth]{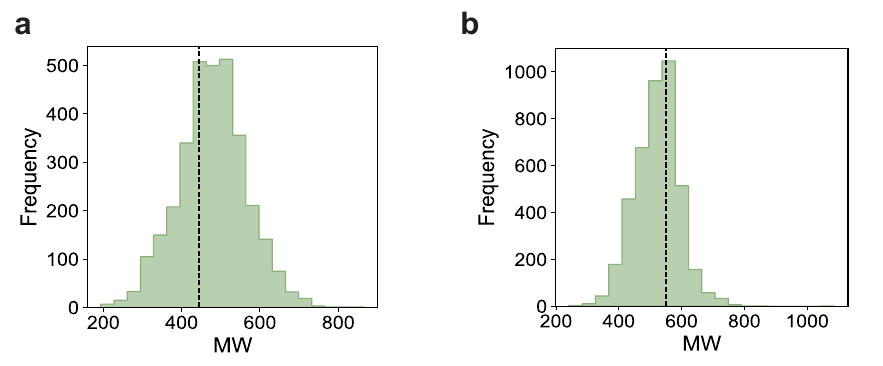} 
    \caption{\textbf{a,b, }Histograms illustrating the molecular weight distributions of the molecular libraries generated by {\ourM} for the MALT1 (\textbf{a}) and CBL-B (\textbf{b}) optimization tasks. The vertical dashed lines represent the property values of the input molecules.}
    \label{fig:mw}
\end{figure}

\clearpage
\subsection*{Supplementary tables}
\begin{table}[!htbp]
\centering
\caption{Property thresholds for training properties.}
\begin{tabular}{ccc}
\toprule
Property & Data source & $\delta$ \\
\midrule
Bioactivity & PubChem & $y_i'$\\
Bioactivity (DRD2) & GeLLMO & 0.2\\
Bioactivity (JNK3) & Synthetic & 0.1\\
Bioactivity (GSK3$\beta$) & Synthetic & 0.1\\
BBBP & Prompt-MolOpt & $y_i' > 0.5$ and $y_i <= 0.5$\\
BBBP & GeLLMO & 0.2\\
HIA & GeLLMO & 0.1\\
Mutag & GeLLMO & 0.1\\
hERG & Prompt-MolOpt & $y_i' <= 0.5$ and $y_i > 0.5$\\
pLogP & GeLLMO & 1\\
water solubility & Prompt-MolOpt & 0.5\\
QED & GeLLMO & 0.1\\
HBA & DrugAssist & 1\\
HBD & DrugAssist & 1\\
\bottomrule
\end{tabular}
\label{tab:delta}
\end{table}
\newpage
\begin{table}[!htbp]
\caption{Instruction templates for bioactivity optimization}
\centering\resizebox{\columnwidth}{!}{
\begin{tabular}{l}
\toprule
\textbf{Template}\\
\midrule
\textbf{Bioactivity}\\
\midrule
Design a molecule by optimizing \{mol1\} to enhance its binding affinity with \{target\_name\}.\\
Given \{target\_name\}, suggest modifications to \{mol1\} to create a new molecule.\\
How can \{mol1\} be modified to better interact with \{target\_name\}?\\
To enhance the binding of \{mol1\} with \{target\_name\}, how can it be optimized?\\
What changes can be made to \{mol1\} to improve its affinity for \{target\_name\}?\\
Develop a derivative of \{mol1\} to enhance interaction with \{target\_name\}.\\
Suggest a modification to \{mol1\} to enhance its binding with \{target\_name\}.\\
What innovative changes can be made to \{mol1\} to optimize its binding with \{target\_name\}?\\
Propose an improvement to \{mol1\} to interact better with \{target\_name\}.\\
What novel modifications could optimize \{mol1\}'s affinity for \{target\_name\}?\\
Given \{target\_name\}, suggest improvements to \{mol1\} to enhance its binding.\\
What structural changes can enhance \{mol1\}'s binding with \{target\_name\}?\\
If the target is \{target\_name\}, what modifications can improve \{mol1\}'s binding?\\
What new features can be added to \{mol1\} to strengthen its interaction with \{target\_name\}?\\
How would you modify \{mol1\} to increase its affinity for the target \{target\_name\}?\\
For \{target\_name\}, what modifications can you make to \{mol1\} to enhance its binding affinity?\\
What creative structural changes can be introduced to \{mol1\} to optimize its binding with \{target\_name\}?\\
\bottomrule
\end{tabular}}
\label{tab:prompt-bioactivity}
\end{table}

\begin{table}[!htbp]
\centering
\caption{Instruction templates for ADMET properties and multi-property combination}
\small
\begin{tabular}{l}
\toprule
\textbf{Template}\\
\midrule
\textbf{BBBP}\\
\midrule
Enhance the BBB penetration capability of \{mol1\}.\\
Boost the ability of \{mol1\} to cross the BBB.\\
Enhance the ability of \{mol1\} to cross the blood-brain barrier.\\
\midrule
\textbf{Water solubility}\\
\midrule
Boost the solubility of \{mol1\} in water.\\
Improve the water solubility of \{mol1\}.\\
Boost the ability of \{mol1\} to dissolve in water.\\
\midrule
\textbf{HBA}\\
\midrule
Improve the hydrogen bond acceptor property of \{mol1\}.\\
Boost the HBA property of \{mol1\}.\\
Enhance the hydrogen bond acceptor (HBA) property of \{mol1\}.\\
\midrule
\textbf{HBD}\\
\midrule
Boost the hydrogen bond donor value of \{mol1\}.\\
Boost the HBD value of \{mol1\}.\\
Improve the hydrogen bond donor property (HBD) of \{mol1\}.\\
\midrule
\textbf{hERG}\\
\midrule
Minimize the hERG inhibition potential of \{mol1\}.\\
Reduce the risk of \{mol1\} affecting the hERG channel.\\
\midrule
\textbf{HIA}\\
\midrule
Increase the intestinal absorption ability of \{mol1\}.\\
Enhance the ability of \{mol1\} to be absorbed in the human intestine.\\
Boost the human intestinal absorption (hIA) of \{mol1\}.\\
\midrule
\textbf{Mutagenicity}\\
\midrule
Reduce the mutagenicity of \{mol1\}.\\
Lower the mutagenicity of \{mol1\}.\\
Decrease the mutagenicity risk of \{mol1\}.\\
\midrule
\textbf{pLogP}\\
\midrule
Increase the LogP of \{mol1\}.\\
Boost the LogP value of \{mol1\}.\\
Enhance the hydrophobicity of \{mol1\}.\\
\midrule
\textbf{QED}\\
\midrule
Increase the QED (quantitative estimate of drug-likeness) of \{mol1\}.\\
Boost the QED score of \{mol1\}.\\
Enhance the drug-likeness of \{mol1\}.\\
\midrule
\textbf{Multi-property}\\
\midrule
\{modify\} the \{properties\} of \{mol1\}.\\
\{modify1\} the \{properties1\} and \{modify2\} the \{properties2\} of \{mol1\}.\\
\bottomrule
\end{tabular}
\label{tab:prompt-admet}
\end{table}
\end{document}